\documentclass[namedreferences]{solarphysics}

\usepackage[hyperref,optionalrh,solaromanenum]{spr-sola-addons} 
\usepackage{graphicx}                    
\usepackage{color}                       

\renewcommand{\vec}[1]{{\mathbfit #1}}

\chardef\us=`\_

\begin{document}

\begin{article}

\begin{opening}

\title{Magnetic Evolution of an Active Region Producing Successive Flares and Confined Eruptions}

\author[addressref={1},email={lopezf@iafe.uba.ar}]{\inits{M. }\fnm{Marcelo }\lnm{L\'opez Fuentes}\orcid{http://orcid.org/0000-0001-8830-4022}}
\author[addressref={1}]{\inits{M. }\fnm{Mariano }\lnm{Poisson}\orcid{http://orcid.org/0000-0002-4300-0954}}
\author[addressref={1}]{\inits{C. H. }\fnm{Cristina H. }\lnm{Mandrini}\orcid{http://orcid.org/0000-0001-9311-678X}}

\runningauthor{L\'opez Fuentes, Poisson, and Mandrini}
\runningtitle{Evolution of an AR Producing Recurrent Events}

\address[id={1}]{Instituto de Astronom\'{\i}a y F\'{\i}sica del Espacio (IAFE), CONICET-UBA, Ciudad Universitaria – Pab. 2, Intendente Güiraldes 2160,
C1428EGA C.A.B.A., Argentina}

\begin{abstract}
We analyze the magnetic evolution of solar active region (AR) NOAA 11476 that, between 9 and 10 May 2012, produced a series of surge-type eruptions accompanied by GOES X-ray class M flares. Using force-free models of the AR coronal structure and observations in several wavelengths, in previous works we studied the detailed evolution of those eruptions, relating them to the characteristic magnetic topology of the AR and reconstructing the involved reconnection scheme. We found that the eruptions were due to the ejection of minifilaments, which were recurrently ejected and reformed at the polarity inversion line of a bipole that emerged in the middle of the positive main AR magnetic polarity. The bipole was observed to rotate for several tens of hours before the events. In this article we analyze, for the full AR and the rotating bipole, the evolution of a series of magnetic parameters computed using the {\it Helioseismic and Magnetic Imager} (HMI) vector magnetograms. We combine this analysis with estimations of the injection of magnetic energy and helicity obtained using the Differential Affine Velocity Estimator for Vector Magnetograms (DAVE4VM) method that determines, from vector magnetograms, the affine velocity field constrained by the induction equation. From our results, we conclude that the bipole rotation was the main driver that provided the magnetic energy and helicity involved in the minifilament destabilizations and ejections. The results also suggest that the observed rotation is probably due to the emergence of a kinked magnetic flux rope with negative writhe helicity.  
\end{abstract}


\end{opening}


\section{Introduction}
\label{s:intro}

Motivated by the potential impact on space weather, one of the main interests behind the study of solar active regions is the ability to determine which characteristics and/or  evolution make them more prone to produce impulsive events like flares and coronal mass ejections (CMEs) \citep[for recent reviews on this topic see, e.g.,][]{toriumi2019,green2018,patsourakos2020}. Among the global characteristics of active regions (ARs), the presence of $\delta$-spots, configurations with quadrupolar or multiple mixed polarities and magnetic complexity in general, interaction with neighbor ARs, and the presence of newly emerged bipoles, have been generally associated with their tendency to produce major energetic events \citep[see, e.g.,][]{sammis2000,vandriel2015,yang2017,toriumi2017}. Particular features have also been identified as precursors of flares and CMEs, such as strong shear along and around polarity inversion lines (PILs) \citep{hagyard1984,schrijver2007,wang2017}, flux cancellation \citep{martens2001,green2011,panesar2017}, and the rotation of magnetic structures and sunspots \citep[][]{zhang2008,ruan2014,vemareddy2016}.\\

Many of the aforementioned observational features are due to the deformation of the magnetic field in the structures emerging from the solar interior to form ARs \citep{pevtsov2014}. These deformations are associated to the presence of free magnetic energy available to be released by field reconnection and reconfiguration once the structures emerge into the solar atmosphere. The magnetic helicity, as a magnetohydrodynamic (MHD) invariant, provides a quantifiable measure of the higher or lesser degree of deformation present in AR magnetic structures and, therefore, of the major or minor tendency of ARs to produce impulsively energetic events. For that reason, there has been a strong interest in recent years to develop methods to compute the injection of magnetic helicity into ARs, both by the emergence process itself and by the deformation of the structures by photospheric motions \citep[see, e.g.,][]{green2002,demoulin2009,thalmann2021}. \\

Magnetic helicity, or more precisely the relative magnetic helicity \citep{berger1984,berger2003}, that we henceforth call $H$, can be defined in complex magnetic configurations and has the important property of being very well preserved in plasmas having a high magnetic Reynolds number (10$^8$\,--\,10$^{15}$), even taking into account the effects of dissipative processes. Using a solar-like eruptive event simulation, \citet{pariat2015} have confirmed that $H$ is quasi-conserved even when non-ideal processes, such as reconnection, act to transform magnetic energy. This now-demonstrated conservation of $H$ is at the base of the initiation of CMEs and their role in ejecting the accumulated coronal helicity \citep{rust1994,low1996,green2002,mandrini2005,priest2016}. Furthermore, the accumulation of helicity in the corona plays a significant role in storing free magnetic energy. \citet{zuccarello2018} used a series of 3D parametric magnetohydrodynamic (MHD) simulations of the formation and eruption of magnetic flux ropes to investigate how different kinds of photospheric boundary flows accumulate $H$ in the corona, and how well $H$-related quantities identify the onset of an eruption \citep[for a different set of simulations see also][]{pariat2017,toriumi2017b}. These authors applied the decomposition of $H$ into two components \citep{berger2003,pariat2015}, i.e. the helicity of the non-potential, or current-carrying, component of the magnetic field, $H_J$, and the mutual helicity between the potential and the current-carrying field, as well as the decomposition of the magnetic energy following \citet{valori2013}. They found that, at the onset of the eruptions, the ratio between the non-potential magnetic helicity, $H_J$, and the total $H$ had the same value for all simulations, suggesting the existence of a threshold in this quantity, but that such a threshold was not observed for quantities related to the magnetic energy. \\

From the observational side, \citet{thalmann2019} used different computational methods to study two ARs with eruptive and confined flare production and confirmed the ability of the current-carrying to total helicity ratio, $H_J/H$, to identify the eruptive potential of an AR. They found, however, that the numerical value of the helicity ratio is not enough to predict the magnitude or type of eruptive flare. They also confirmed the previous result \citep{labonte2007} that the ratio of total helicity to the square of the magnetic flux ($H/\phi^{2}$) is not a good indicator of eruptivity. \citet{gupta2021} found, based on their study of 10 ARs producing large flares, precise limits of $H_J/H$ and other quantities to segregate ARs producing eruptive and confined flares \citep[see also][]{tziotziou2012,tziotziou2014}. \\

There are basically two families of methods to compute magnetic energy and helicity in observed ARs: finite volume methods \citep[see][]{valori2016}, based on computing these magnitudes from models of the AR magnetic structure extrapolated from photospheric magnetograms into a finite coronal volume; and flux integration methods \citep[see, e.g.,][]{pariat2005,liu2012}, based on the injection and accumulation of magnetic energy and helicity computed from the evolution of observed photospheric magnetograms along a given time. \citet{thalmann2021} compared some of these methods by applying them to a single case of an AR evolution associated to an X-class eruptive flare. They found a reasonable correspondence between the results obtained with the different methods but concluded that, thus far, a full description of the evolution of an event may only be achieved by the combined use of these methods. Similarly, \citet{liokati2022}, based on their analysis of the emergence process of CME-producing ARs, suggested that any study of the eruptive potential of ARs should consider magnetic helicity together with magnetic energy. \citet{liu2023} applied a superposed epoch analysis method to a set of eruptive and non-eruptive flares from different ARs, by analyzing the magnetic energy and helicity obtained with a finite volume method based on non-linear force-free models, and a flux integration method based on the photospheric velocity field computed using the differential-affine velocity procedure \citep[DAVE,][]{schuck2008,welsch2009}. They found that while eruptive events tend to show a marked decrease in magnetic energy and helicity during the events, non-eruptive cases do not exhibit variations larger than the associated errors. They also reported that magnetic energy and helicity levels replenished back to pre-event values after an average of 12 h after the events. Curiously, in the eruptive cases they also observed a sudden decrease of the photospheric twist of the ARs before the events, followed by an increase right after them. This behavior suggests some kind of bounce back mechanism at work during the eruptions that would require further study in the future. \\

The widespread availability of vector magnetogram data at a reasonable cadence brought the possibility of computing on a regular basis series of magnetic parameters to characterize the major or minor tendency of ARs to produce flares and ejections \citep[see, e.g.,][]{leka2007}. This was implemented as part of the Space-Weather HMI AR Patches (SHARP) pipeline developed by \citet{bobra2014}, which includes AR global magnetic parameters computed from vector magnetograms obtained with the {\it Helioseismic Magnetograph Imager} \citep[HMI,][]{scherrer2012} on board the {\it Solar Dynamics Observatory} (SDO).  This database has been later extended to include data from the {\it Michelson-Doppler Imager} \citep[MDI,][]{scherrer1995} on board the {\it Solar and Heliospheric Observatory} (SOHO) \citep{bobra2021}. Although due to the stochastic nature of flaring and eruptive events no magnetic parameter could be identified as a single precursor of an imminent event \citep[see ][]{kontogiannis2023}, growing sets of parameters are being considered in large scale statistical studies focused on flare and CME forecasting \citep{leka2019a,leka2019b}, including those applying machine learning methods \citep[see, e.g.,][]{bobra2015,bobra2016,nishizuka2017,georgoulis2021}. \\

In this article, we study the magnetic evolution of AR NOAA 11476, which in approximately 16 h between 9 and 10 May 2012, produced three confined ejections of the surge type due to minifilement eruptions associated to M-class flares. In previous works \citep{lopezfuentes2018,poisson2020}, we studied the topology and evolution of the magnetic configuration that produced the first two of these events. The third event was studied by other authors \citep{yang2018}. Here, we focus on the evolution of the photospheric magnetic field along several days around the time of the events to analyze what are the characteristics that define the non-potentiality of the AR that provided the free magnetic energy released during the flares and eruptions. For that, we use a series of photospheric magnetic parameters obtained from SHARP data, both for the full AR and a partial subregion of interest. We also compute the injected magnetic energy (through the Poynting flux) and magnetic helicity using the DAVE for vector-magnetograms \citep[DAVE4VM,][]{schuck2008} tool.\\

In Section~\ref{s:ar} we briefly summarize the global evolution and characteristics of the AR and the eruptions, in Section~\ref{s:methods} we describe the data and methods used for the analysis, in Section~\ref{s:results} we present our results, and we discuss and conclude in Section~\ref{s:conclusion}.\\


\section{The Evolution of AR 11476}
\label{s:ar} 

\subsection{Photospheric Evolution}
\label{s:evol}

\begin{figure} 
\centerline{\includegraphics[width=0.8\textwidth,clip=]{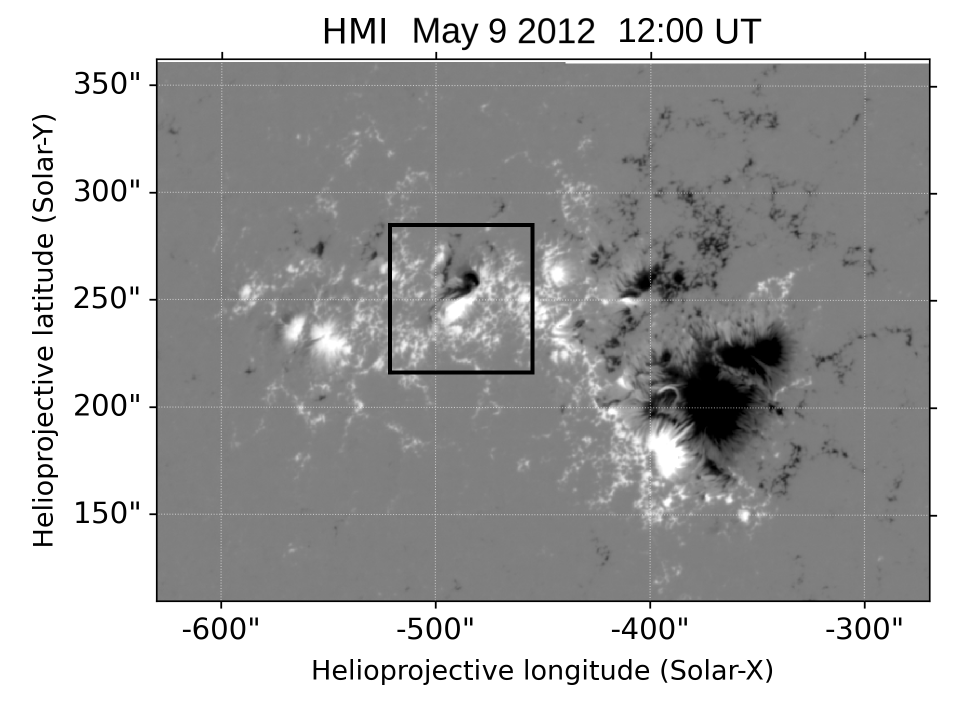}}
\caption{SDO/HMI line-of-sight magnetogram of AR 11476. The date and universal time of the data are provided at the top of the panel. Black and white areas correspond, respectively, to negative and positive magnetic fields. Magnetic field values are saturated in $\pm$ 1000 G. The black square indicates the location of the bipolar structure whose rotation (see Figure~\ref{fig:bipole}) was the source of the injection of magnetic energy and helicity that eventually produced three confined eruptive events. The units in the axes are arcsec in the plane of the sky from the solar disk center.}
   \label{fig:ar}
\end{figure}

\begin{figure} 
\centerline{\includegraphics[width=0.9\textwidth,clip=]{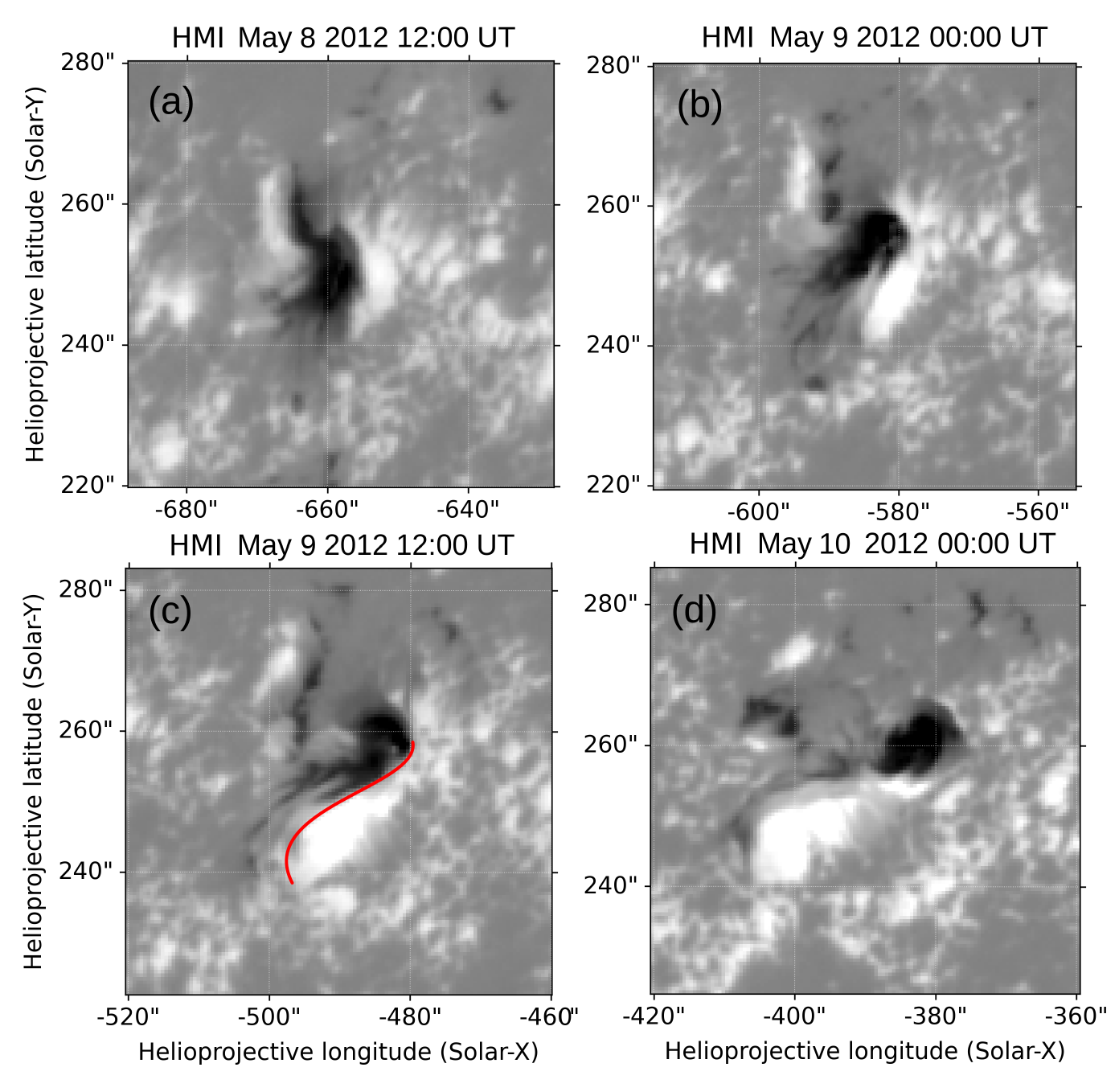}}
\caption{The panels show the evolution of the rotating bipole that emerged in the middle of the main positive polarity of AR 11476. The area covered by the panels approximately coincides with the square box plotted in Figure~\ref{fig:ar}. The dates and universal times of the magnetograms are provided at the top of the panels. Axes units are in arcsec in the plane of the sky with respect to the solar disk center. The red curve in panel c indicates the approximate location of the bipole PIL, where mini-filaments were successively ejected and reformed during a lapse of approximately 16 h between May 9 and 10, 2012. The conventions for the horizontal and vertical axes, as well as that for the magnetic field, are the same as in Figure~\ref{fig:ar}.}
   \label{fig:bipole}
\end{figure}

AR NOAA 11476 appeared on the eastern solar limb on May 4, 2012. By May 7 (around the time when our analysis begins) and successive days, the AR, as observed in line of sight (LOS) HMI magnetograms, consisted of a generally bipolar configuration formed by a relatively compact leading negative polarity followed by a more extended and fragmented positive polarity. Figure~\ref{fig:ar} shows the photospheric magnetic field of the AR around 12:00 UT on May 9, 2012, minutes before the first of the surge-type eruptions mentioned in Section~\ref{s:intro}. From the beginning of the analyzed evolution, a smaller bipolar structure is observed in the middle of the main positive polarity, as highlighted with a black square in the panel of Figure~\ref{fig:ar}. As described in \citet{lopezfuentes2018} this bipole is observed to rotate by almost 180$^\circ$ between the beginning of May 8 and the end of May 10. Figure~\ref{fig:bipole} shows four snapshots of the bipole magnetic field during the 36 h period of faster rotation. The area covered by the panels approximately coincides with the black square area of Figure~\ref{fig:ar}. Times are indicated at the top of the panels. In \citet{lopezfuentes2018} and \citet{poisson2020} we concluded that this rotation was the main source of free magnetic energy and helicity injected into the AR to produce the studied eruptive events. Notice that at the north-east and very close to the bipole there is another smaller bipole, which is also observed to rotate (see its evolution in the panels of Figure~\ref{fig:bipole}) indicating that all that area of the AR could be the location of magnetic helicity injection via rotational motions or the emergence of deformed magnetic structures. Whenever we refer to the "bipole" in the rest of the paper we will be referring to the larger bipolar region. 

\begin{figure} 
\centerline{\includegraphics[width=0.8\textwidth,clip=]{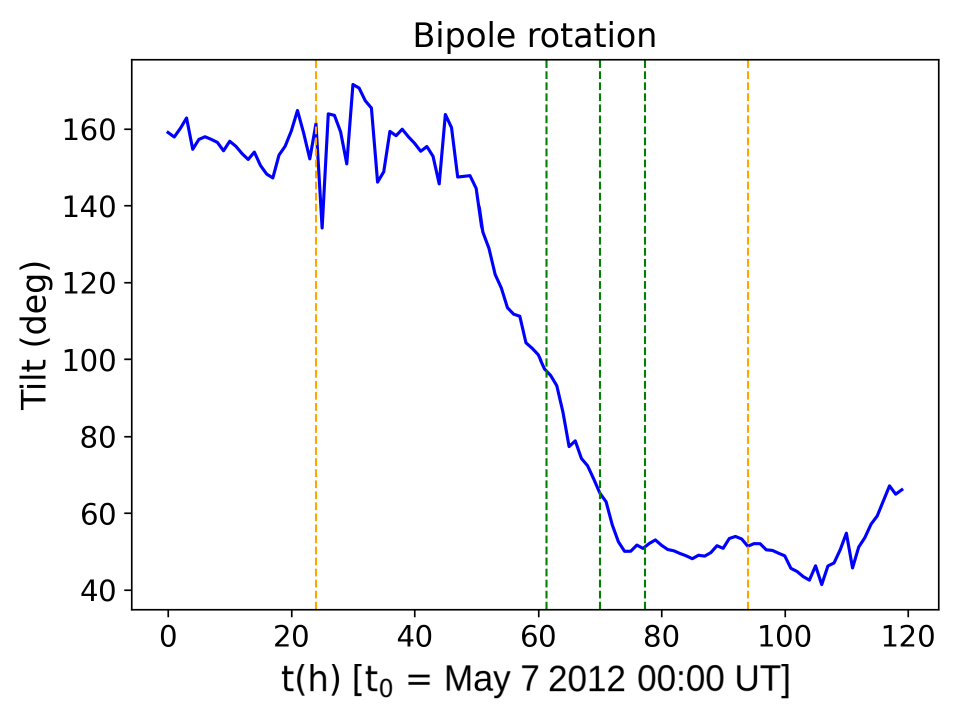}}
\caption{Evolution of the tilt angle of the bipole computed as the angle that the vector joining the main magnetic polarity barycenters (from positive to negative) form with the solar equator. The tilt angle is defined positive in the counter-clockwise direction from the east-west line parallel to the solar equator. Magnetic fields considered in the computation are above (below) 500 G (-500 G). Time is in hours from May 7, 2012, 00:00 UT. The vertical dashed yellow lines indicate the times between which the bipole rotation was identified in \citet{lopezfuentes2018} and \citet{poisson2020}. The vertical dashed green lines indicate the times when the eruptive events described in Section~\ref{s:eruptions} occurred.
}
   \label{fig:rotation}
\end{figure}

To quantify the observed rotation of the bipole, in Figure~\ref{fig:rotation} we show the evolution of its tilt, defined as the angle that the vector joining the positive and negative magnetic polarity barycenters forms with the direction of the solar equator. The barycenters are computed as the magnetically weighted geometrical centers of the positive and negative polarities as defined in \citet{lopezfuentes2000} \citep[see also][]{poisson2015}. The computation is done over solar radialized magnetograms with a cadence of 1 h and considering magnetic fields above (below) 500 G (-500 G). The vector goes from the center of positive polarity to the center of negative polarity. The tilt angle is defined as positive in the counter-clockwise direction starting from the line parallel to the solar equator in the usual positive x-axis direction. In the plot, time is provided in hours from the time of the initial analyzed magnetogram, on May 7, 2012, at 00:00 UT. The yellow dashed vertical lines indicate the times between which the main rotation of the bipole was identified in our previous works \citep{lopezfuentes2018,poisson2020} by visual inspection of a series of observed magnetograms, while the green dashed vertical lines correspond to the times when the eruptive events described in Section~\ref{s:eruptions} occurred. As seen in Figure~\ref{fig:rotation}, the bipole's tilt angle evolution can be separated in a gradually, although noisy, negative rotation between t$\approx$24 h and t$\approx$45 h and a much faster rotation from then up to t$\approx$75 h, just before the last eruptive event occurs.

\subsection{Eruptive Events}
\label{s:eruptions}

\begin{figure} 
\centerline{\includegraphics[width=0.9\textwidth,clip=]{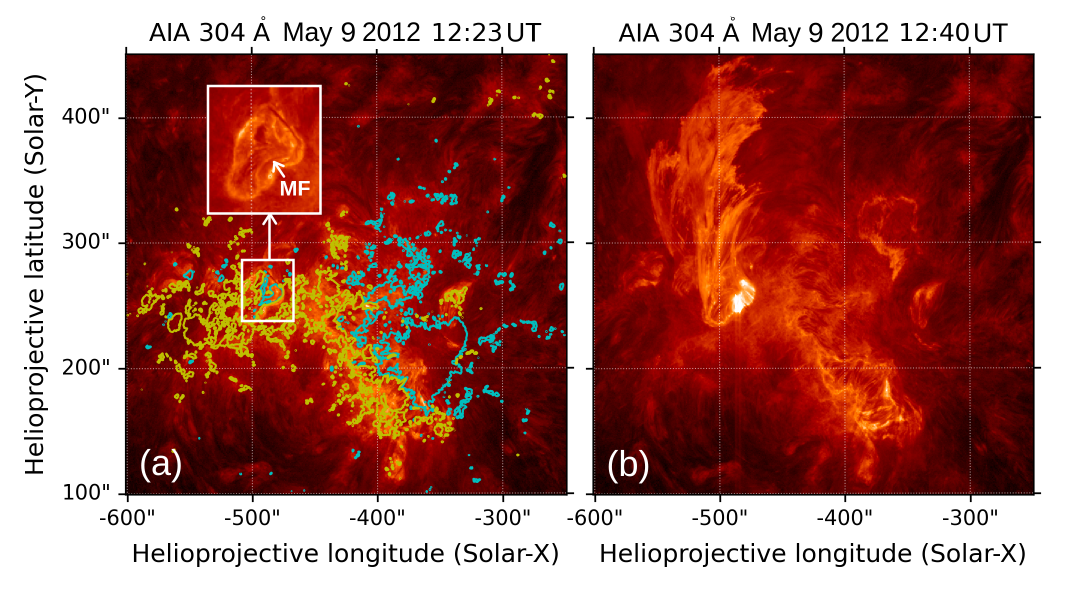}}
\caption{SDO/AIA images in the 304 $\mathrm{\AA}$ channel of AR 11476 around the time of the first mini-filament eruption. Date and universal times are provided at the top of the panels. Panel a corresponds to a time just before the eruption. Green and yellow contours correspond to $\pm$200 G levels of a nearly co-temporal HMI LOS magnetogram. The inset shows a zoom of the area of the bipole where the mini-filament location is indicated with a white arrow. Panel b shows the surge-type eruption close to the time of its maximum apparent extension, some 17 min after its initiation.}
   \label{fig:eruption}
\end{figure}

Between mid day May 9 and the first hours of May 10, three ejections of the surge-type, as observed in different wavelengths, were produced in the area around the polarity inversion line (PIL) of the larger bipole. The first event occurred at 12:23 UT on May 9, 2012, and was associated to an M4.7 flare \citep[see][]{lopezfuentes2018}; the second one, consisting actually in a two-step ejection, started at 21:01 UT on the same day and was associated to a M4.1 flare \citep{poisson2020}; and the third event occurred at 4:18 UT on May 10 and was associated to a M5.7 flare \citep{yang2018}. The flare emissions are clearly identifiable in the X-ray light-curve from the Geostationary Operational Environmental Satellite \citep[GOES, see Figure 1 of][]{poisson2020}. The origin of the ejections was the successive formation and eruption of mini-filaments located along the bipole PIL. The red line in Figure~\ref{fig:bipole}, panel c, indicates the location of the PIL where the mini-filaments were formed.  The observed photospheric evolution indicated that the whole process was due to the combined effect of the rotation of the photospheric magnetic field structure (injecting magnetic helicity and free energy) and flux cancellation at the PIL. The combination of rotation and magnetic flux cancellation was analyzed in \citet[][see their Figure 4]{poisson2020}.\\

Figure~\ref{fig:eruption}, panel a, shows an image of the AR observed in the 304 $\mathrm{\AA}$ channel of the Atmospheric Imaging Assembly \citep[AIA,][]{lemen2012} right before the start of the first event. A zoomed inset of the area around the bipole is added; in this inset the mini-filament (identified as MF) location is indicated with a white arrow. Overlaid yellow and cyan contours correspond to $\pm$200 G levels of the approximately co-temporal HMI LOS magnetogram. The topological analysis done using force-free models of the AR coronal magnetic structure, in combination with observations in different wavelengths, allowed us to reconstruct the evolution of the eruptions, which consisted of two reconnection processes, one taking place below the mini-filament, that pushed the structure upwards, and another above the mini-filament that reconnected the front of the ascending structure with overlying longer magnetic field lines belonging to the larger structure of the AR. These larger loops were then filled with mini-filament material as illustrated in the AIA 304 image of Figure~\ref{fig:eruption}b; this panel corresponds to the time of maximum apparent extension of the surge. The whole process is thoroughly described in our previous works \citep{lopezfuentes2018,poisson2020}.  


\section{Data and Method Descriptions}
\label{s:methods}

\subsection{SHARP Photospheric Magnetic Parameters}
\label{s:sharp}

To study the characteristics of the magnetic evolution of AR 11476 that led to the production of the three eruptive events described in Section~\ref{s:eruptions}, we use a series of magnetic parameters included in the database of the Space-weather HMI Active Region Patches \citep[SHARP,][]{bobra2014}. The SHARP package provides HMI data products in the form of Flexible Image Transport System (FITS) files containing LOS and vector magnetograms, Doppler velocity maps, and continuum intensity images of ARs identified using an automatic tracking procedure. Each magnetic concentration automatically identified as an AR is called an HMI Active Region Patch (HARP). 
Since the procedure identifies more field concentrations than those corresponding to NOAA ARs, HARPs follow their own numeration. Each NOAA AR has then its corresponding HARP identification number, which in the case of AR 11476 is 1638. HARP datasets are produced at a rate of one every 12 min, corresponding to the HMI vector magnetogram cadence. \\

\begin{table}
\caption{Names and mathematical expressions of the magnetic parameters used in the analysis of the evolution of AR 11476 \citep[as provided by][]{bobra2014}.}
\label{tbl:params}
\begin{tabular}{cc}    
\hline
Parameter & Expression \\
\hline
Horizontal gradient of horizontal field &
$\overline{|\nabla B_h|} = \frac{1}{N} \sum{\sqrt{\left(\frac{\partial B_h}{\partial x}\right)^2+ \left(\frac{\partial B_h}{\partial y}\right)^2}}$ \\
\\
Horizontal gradient of vertical field &
$\overline{|\nabla B_z|} = \frac{1}{N} \sum{\sqrt{\left(\frac{\partial B_z}{\partial x}\right)^2+ \left(\frac{\partial B_z}{\partial y}\right)^2}}$ \\
\\
Total unsigned magnetic flux &
$\Phi = \sum{|B_z| dA}$\\
\\
Total unsigned vertical current &
$J_{z_{total}} = \sum{|J_z| dA}$\\
\\
Mean excess of magnetic energy density &
$\overline{\rho} \propto \frac{1}{N} \sum \left(\vec{B}^{Obs}-\vec{B}^{Pot}\right)^{2}$\\
\\
Total magnetic free energy density &
$\rho_{tot} \propto \sum \left(\vec{B} ^{Obs}-\vec{B} ^{Pot}\right)^{2}~dA$\\
\\
Mean twist parameter $\alpha$ &
$\alpha_{total} \propto \frac{\sum J_{z} \cdot B_{z}}{\sum B_{z}^{2}}$\\
\\
Total unsigned current helicity &
$H_{c_{total}} \propto \sum |B_z \cdot J_z|$\\ 
\hline
\end{tabular}
\end{table}

Global magnetic parameters computed over each HARP are included in the FITS header accompanying the corresponding data array. These include 16 indices associated with characteristics such as the total magnetic flux, the spatial gradients of the field, the vertical current density and current helicity, and proxies for the free magnetic energy. These are parameters which have been associated with the level of non-potentiality of the AR and regarded as possible indicators of flare productivity \citep{leka2007}. The parameters are computed only on magnetogram pixels accomplishing a threshold of confidence defined by the quality of the disambiguation of the transverse magnetic field. The full list of parameters and their mathematical expressions are provided in \citet{bobra2014}. \\

\begin{figure} 
\centerline{\includegraphics[width=\textwidth,clip=]{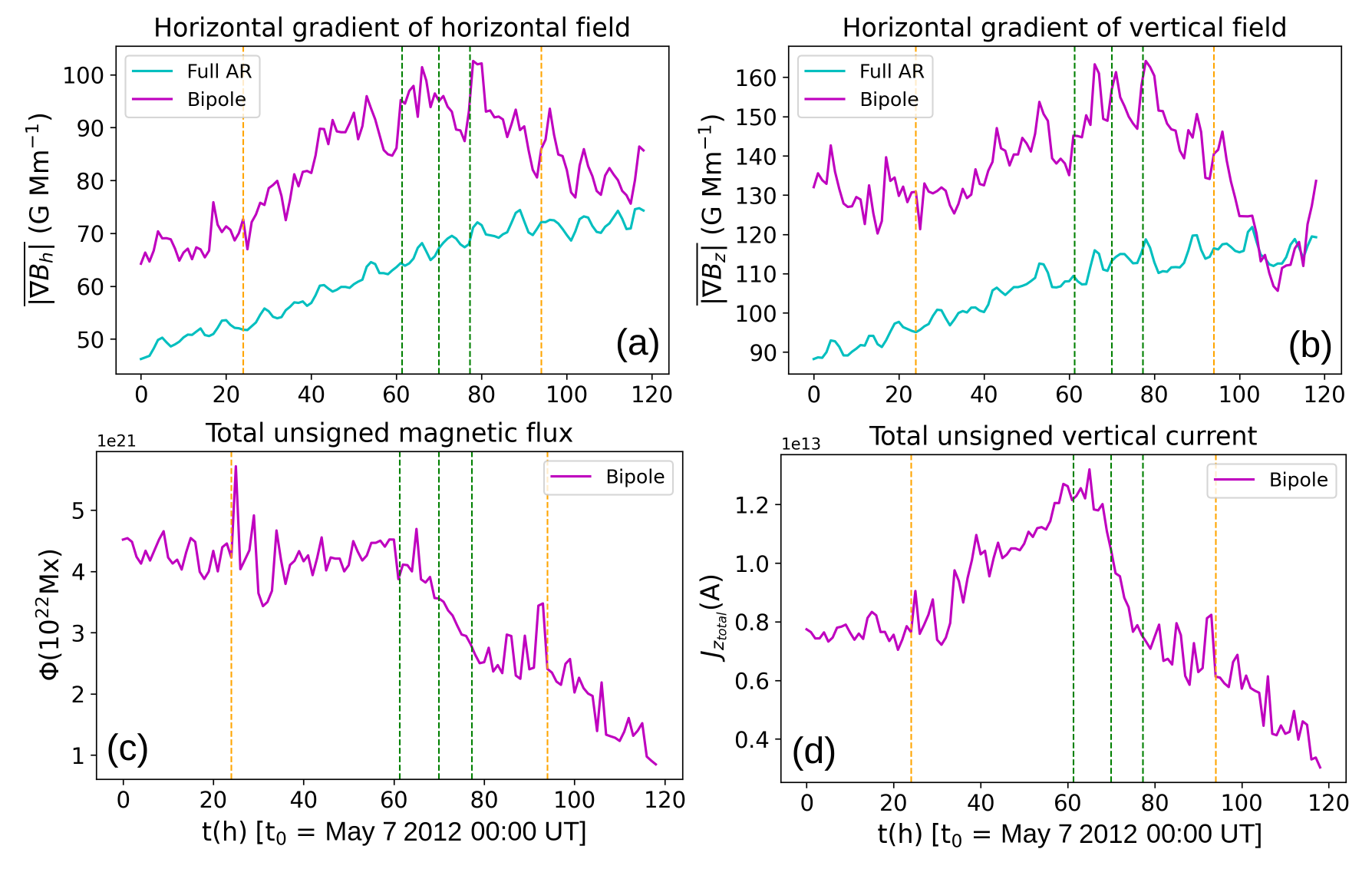}}
\caption{Evolution of the four  magnetic parameters computed from HMI vector magnetograms for the full AR 11476 (cyan curves where they are represented) and the area corresponding to the bipole (in magenta). The names of the parameters appear at the top of the panels. Time is measured in hours from the date and universal time of the first analyzed magnetogram, as indicated in the label of the horizontal axes ($t_0 =$ May 7 2012 00:00 UT). The vertical dashed yellow lines indicate the approximate beginning and end of the period along which the bipole was observed to rotate. The vertical dashed green lines indicate the times of the confined eruptions.} 
   \label{fig:sharp1}
\end{figure}

\begin{figure} 
\centerline{\includegraphics[width=\textwidth,clip=]{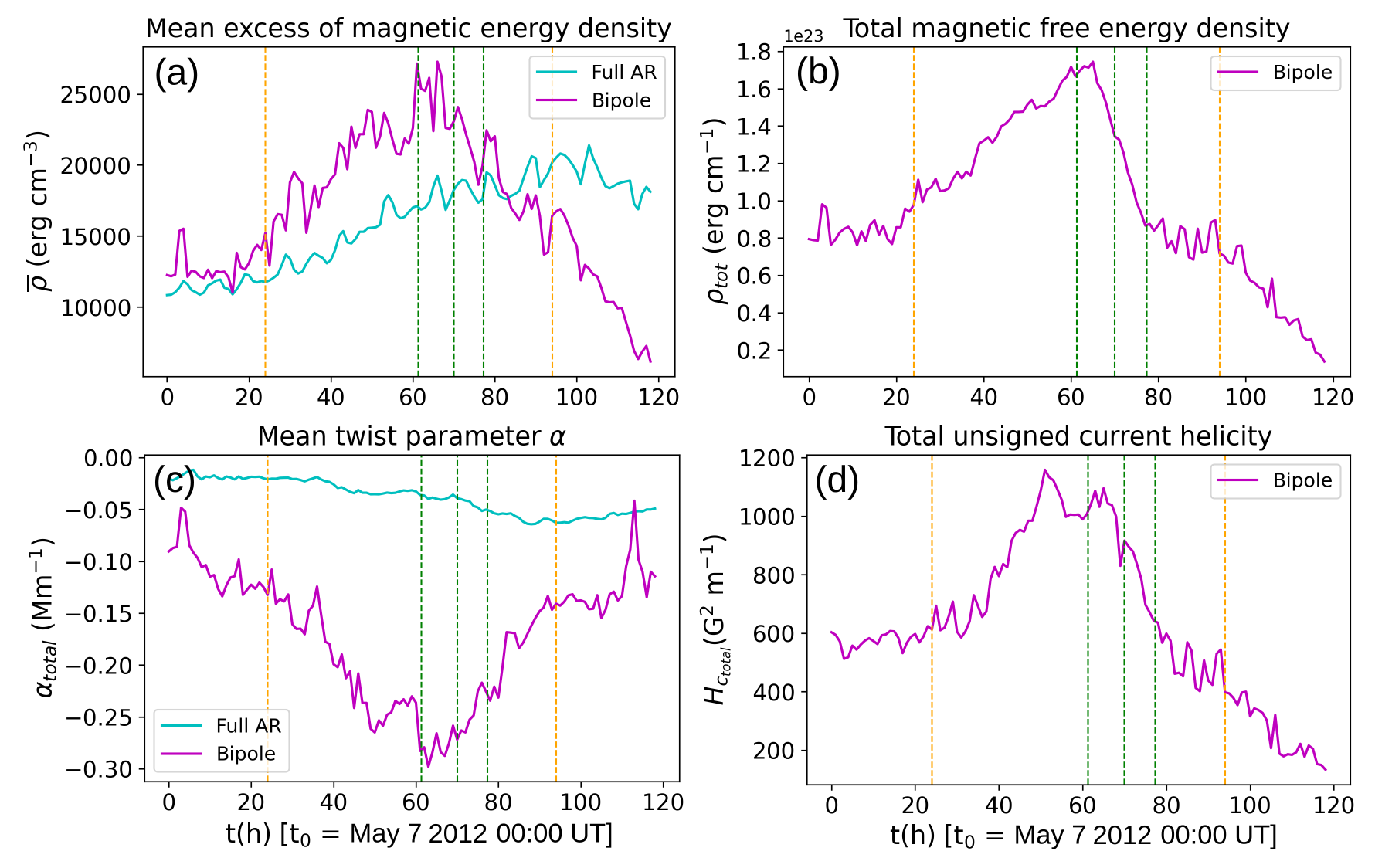}}
\caption{Evolution of the four magnetic parameters computed from HMI vector magnetograms for the full AR 11476 (cyan curves where they are represented) and the area corresponding to the bipole (in magenta). The names of the parameters are shown at the top of the panels. Time is measured in hours from the date and universal time of the first analyzed magnetogram, as indicated in the label of the horizontal axes ($t_0 =$ May 7 2012 00:00 UT). The vertical dashed yellow lines indicate the approximate beginning and end of the period along which the bipole was observed to rotate. The vertical dashed green lines indicate the times of the confined eruptions.}
   \label{fig:sharp2}
\end{figure}

In the present analysis, we use the magnetic parameters from the SHARP datasets of AR 11476 for the complete five days between May 7 and 11, 2012. The SHARP team produces near-real-time versions of the data and a more carefully processed definitive version at later times. Here, we use the definitive version of the processed data. Since we are interested in both, the magnetic evolution of the full AR, but also of the area containing the bipole in an isolated way, we make our own computation of the parameters covering only that area. To select the area of interest around the bipole we use an automatic optimization procedure that encases, for each LOS magnetogram, the maximum magnetic flux that simultaneously accomplishes flux balance within the selected region. In Section~\ref{s:results} we describe and analyze the evolutions of eight of the 16 SHARP magnetic parameters used in our analysis with a cadence of one magnetogram per hour (see Figures~\ref{fig:sharp1} and~\ref{fig:sharp2}). The selection of these eight parameters, out of the 16 analyzed, is based on the fact that the rest of the parameters correspond to similar proxies of the AR magnetic characteristics, showing the same kind of evolution as the selected eight. In Table~\ref{tbl:params} we reproduce the mathematical expressions of the eight parameters shown in the figures, as provided by \citet{bobra2014}.

\subsection{Magnetic Energy and Helicity Injection}
\label{s:helicity}

As part of the analysis of the magnetic evolution of AR 11476, we compute the injection of magnetic energy and helicity into the AR over the time indicated in Section~\ref{s:sharp}, May 7-11, 2012. Following \citet{liu2012}, the magnetic energy input through the AR photospheric boundary, $S$, is computed using \citep{kusano2002}

\begin{equation}
\label{eq:energy}
\left. \frac{dE}{dt} \right|_{S}=\frac{1}{4\pi} \int_{S} B_{t}^2 V_{\bot n} dS - \frac{1}{4\pi} \int_{S}\left(\vec{B}_{t} \cdot \vec{V}_{\bot t}\right)B_{n} dS,
\end{equation}

\noindent where $B_{t}$ and $B_{n}$ are, respectively, the tangent and normal components of the magnetic field with respect to the photosphere obtained from deprojected vector magnetograms, and  $V_{\bot t}$ and $V_{\bot n}$ are the photospheric tangent and normal components of the velocity projected in the direction perpendicular to the magnetic field. Written in this way, the first term of the right hand side of Equation~\ref{eq:energy} corresponds to the injection by flux emergence, while the second term corresponds to the injection by photospheric shear (tangent) motions. Similarly, the magnetic helicity injection can be separated into emergence and shear terms as follows \citep{berger1984},

\begin{equation}
\label{eq:helicity}
\left.\frac{dH}{dt}\right|_{S} = 2 \int_{S}\left(\vec{A}_{p} \cdot \vec{B}_{t}\right)V_{\bot n}dS - 2 \int_{S}\left(\vec{A}_{p}\cdot\vec{V}_{\bot t}\right)B_{n} dS.
\end{equation}

In the above equation, $H$ is what we simply refer as magnetic helicity, but it is in fact the relative magnetic helicity (the difference between the actual magnetic helicity and the magnetic helicity of the potential field having the same normal component as the actual magnetic field at the photospheric boundary). In Equation~\ref{eq:helicity}, $\vec{A}_{p}$ is the vector potential of the potential field, related to the normal component of the magnetic field through $\nabla \times \vec{A}_{p} \cdot \hat{\vec{n}} = B_{n}$, where $\hat{\vec{n}}$ is the normal unit vector to the photosphere.  $\vec{A}_{p}$ also satisfies the Coulomb gauge $\nabla \cdot \vec{A}_{p} = 0$ and the boundary condition $\vec{A}_{p} \cdot \hat{\vec{n}} = 0$. With these conditions, in Cartesian coordinates $\vec{A}_{p}$ can be expressed as \citep{pariat2005}

\begin{equation}
\label{eq:potential}
\vec{A}_{p}(\vec{x}) = \frac{1}{2\pi} \hat{\vec{n}} \times \int_S B_{n}(\vec{x}') \frac{\vec{x}-\vec{x}'}{|\vec{x}-\vec{x}'|^{2}} dS',
\end{equation}

\noindent where $\vec{x}$ and $\vec{x}'$ correspond to two vector positions on the photospheric magnetogram. By replacing Equation~\ref{eq:potential} in Equation~\ref{eq:helicity} one obtains a full expression for the magnetic helicity injection which can be computed directly from deprojected vector magnetograms. The only parameters still to be determined in Equations~\ref{eq:energy} and~\ref{eq:helicity} are the normal and tangential components of the velocity, $V_{\bot t}$ and $V_{\bot n}$. These are obtained using the Differential-Affine Velocity for Vector-magnetograms procedure \citep[DAVE4VM,][]{schuck2008}, which estimates the velocity by a Pearson correlation and Spearman rank order optimization of the relation between the magnetic and velocity fields given by the induction equation,

\begin{equation}
\label{eq:induction}
\frac{\partial B_z}{\partial t} + \nabla_t \cdot \left(V_n \vec{B}_t - \vec{V}_t B_n \right) = 0,
\end{equation}
 
\noindent where $\nabla_t$ corresponds to photospheric tangential derivatives. Finally, to keep only the relevant velocity components, $V_{\bot t}$ and $V_{\bot n}$, the following relation is used

\begin{equation}
\label{eq:vperp}
\vec{V}_\bot = \vec{V} - \frac{\vec{V}\cdot\vec{B}}{B^2} \vec{B},
\end{equation}
 
\noindent where $\vec{V}$ is the velocity computed using the DAVE4VM method.
 
Due to the temporal dependence of Equation~\ref{eq:induction}, each velocity computation requires two successive magnetograms to be performed. In order to have the same temporal resolution of one datapoint per hour as for the SHARP parameters (see Section~\ref{s:sharp}), we use one pair of successive magnetograms, separated by 12 minutes (the maximum HMI cadence), per hour. The first magnetogram of each pair corresponds to that from which the SHARP magnetic parameters were obtained, as described in Section~\ref{s:sharp}.  

\begin{figure} 
\centerline{\includegraphics[width=\textwidth,clip=]{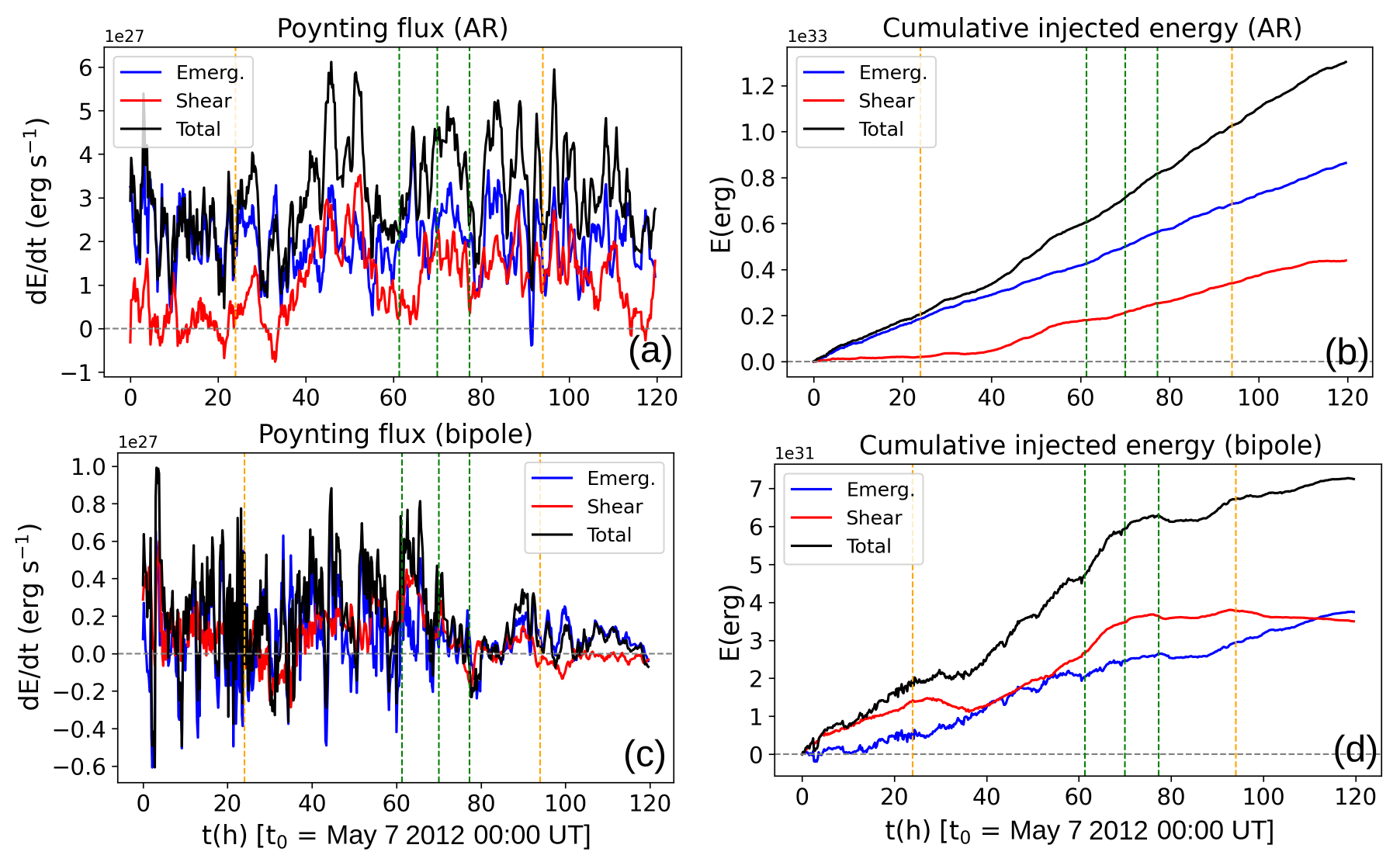}}
\caption{Evolution of the magnetic energy injection for AR 11476. (a): Injection rate computed at a cadence of 1 per hour from two successive HMI vector magnetograms for the full AR area. The curve has been smoothed with a 5-point running average for clarity. (b): Accumulated integration of the rate plotted in panel a. (c): Same as panel a for the magnetogram area containing only the bipole. (d): Accumulated integration of the rate plotted in panel c.}
   \label{fig:poynting}
\end{figure}

In Figures~\ref{fig:poynting} and~\ref{fig:helicity} we show the results of applying the procedure just described to the HMI vector magnetograms considered for the evolution of AR 11476. Figure~\ref{fig:poynting} corresponds to the magnetic energy injection, where panels a and b correspond to the full AR, while panels c and d consider only the area of the bipole. Panels on the left (a and c) correspond to the total injection rates in erg s$^{-1}$ for the area considered, while panels on the right (b and d) show the accumulated sum of the injected energy, assuming an approximately constant rate along each integrated hour interval, corresponding to the value computed at the beginning of the interval. Since the hourly injection rates are very noisy, in panels a and c the corresponding curves were smoothed using 5-point running average. Similarly, Figure~\ref{fig:helicity} follows the same conventions as Figure~\ref{fig:poynting} but for the magnetic helicity injection rate and the corresponding accumulated sums.


\section{Analysis of the results}
\label{s:results}

In Figures~\ref{fig:sharp1} and~\ref{fig:sharp2} we show the evolution of the magnetic parameters described in Section~\ref{s:sharp}, whose mathematical expressions are provided in Table~\ref{tbl:params}. The names of the parameters are provided at the top of each panel. These selected magnetic parameters, which are directly related to the non-potentiality of the photospheric magnetic field, are the ones whose variations are most clearly correlated with the evolution of the bipole rotation. As in Figure~\ref{fig:rotation}, in the panels of  Figures~\ref{fig:sharp1} and~\ref{fig:sharp2} we have added dashed vertical lines indicating, in yellow color the times between which the bipole was observed to rotate and in green color the times when the eruptive events described in Section~\ref{s:eruptions} occurred. In both figures, the magenta curves correspond to the evolution of the parameters computed in the bipole area, described in Section~\ref{s:sharp}, while the cyan curves, when included, correspond to the computation of the parameters on the full AR area. The latter curves are only included in the panels corresponding to parameters that are intensive quantities, like mean values or densities, for which the computation over the bipole and the AR areas have comparable orders of magnitude (panels a and b in Figure~\ref{fig:sharp1} and panels a and c in Figure~\ref{fig:sharp2}). Extensive parameters, like total sums over the analyzed areas, are up to a couple of orders of magnitude different for the bipole and the full AR, so they cannot be represented using comparable scales (panels c and d in Figure~\ref{fig:sharp1} and panels b and d in Figure~\ref{fig:sharp2}). As in Figure~\ref{fig:rotation}, in the time axes, we use hours after the observation date of the first considered magnetogram (very near 00:00 UT of May 7, 2023, as indicated in the $x$ axis legends of the lower panels).

\begin{figure} 
\centerline{\includegraphics[width=\textwidth,clip=]{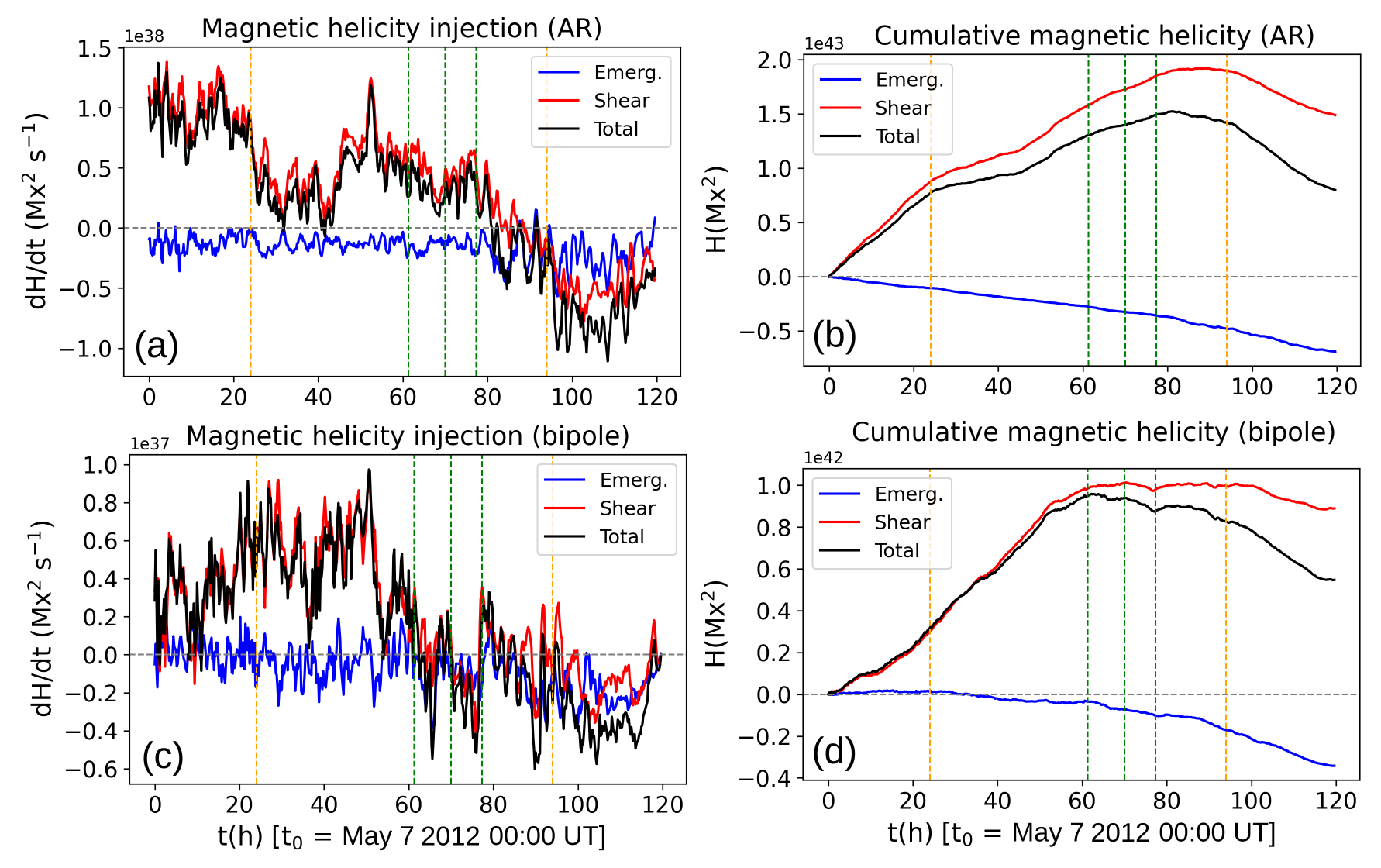}}
\caption{Evolution of the magnetic helicity injection for AR 11476. (a): Injection rate computed at a cadence of one per hour from two successive HMI vector magnetograms for the full AR area. The curve has been smoothed with a 5-point running average for clarity. (b): Accumulated integration of the rate plotted in panel a. (c): Same as panel a for the magnetogram area containing only the bipole. (d): Accumulated integration of the rate plotted in panel c.}
   \label{fig:helicity}
\end{figure}

Panels a and b of Figure~\ref{fig:sharp1} show, respectively, the evolution of the mean horizontal gradients of the horizontal and vertical components of the photospheric magnetic field. Both parameters have a similar evolution. Over the fluctuations, a clear increase of the average values of the parameters begins in coincidence with the start time of the observed bipole rotation (indicated in the panels with the first dashed vertical yellow line). This tendency continues up to the time when the first eruption occurs, keeping an approximately constant mean during the events (green vertical dashed lines). Both parameters are observed to decrease after the events. The relation between the field horizontal gradients and the bipole rotation is basically due to the shear that the rotation produces on the photospheric field. Comparatively, the cyan curves, corresponding to the full AR, show a smoother increase, that seems to stop after the bipole rotation ceases, suggesting that part of the AR field gradients increase may be due to this contribution. 

Regarding the magnetic flux, represented in Figure~\ref{fig:sharp1}, panel c, it maintains an approximately constant average during the first part of the bipole rotation, up to the time of the first eruption, when it starts to continuously decrease. This magnetic flux decrease was identified in our previous studies \citep{lopezfuentes2018,poisson2020}, in which the analysis suggested that flux cancellation was one of the main triggers of the magnetic reconnection process that initiated the mini-filament eruptions. 

The remaining parameter of Figure~\ref{fig:sharp1}, presented in panel d, corresponds to the total unsigned vertical current, another magnitude identified with the non-potentiality of ARs \citep{leka2007}. It shows a dramatically clear correlation with the rotation of the bipole and the occurrence of the events. Once again, there is a sustained average increase up to the time of the first eruption, followed by an abrupt decrease during the events that continues afterwards at a 
lower rate.  

The dramatic variations observed in the unsigned vertical current measured over the bipole area are also present in all the parameters of Figure~\ref{fig:sharp2}, where we show the evolution of the mean excess of magnetic energy density (panel a), the total magnetic free energy density (panel b), the mean twist parameter $\alpha$ (panel c), and the total unsigned current helicity (panel d). In the panels corresponding to intensive parameters (a and c), we also include the evolution curves of those computed for the full AR (cyan curves). Except for the mean $\alpha$, the rest of the parameters shown in Figures~\ref{fig:sharp1} and~\ref{fig:sharp2} are positive by definition. The negative sign of the mean $\alpha$ is consistent with the expected tendency of northern hemisphere ARs to have negative magnetic helicity sign, according to the known hemispheric rule \citep[see, e.g.,][]{pevtsov2014}.

As in the case of the intensive parameters of Figure~\ref{fig:sharp1} (panels a and b) in Figure~\ref{fig:sharp2}, panels a and c, the mean excess of magnetic energy density and the mean twist for the full AR (cyan curves) show a smooth increase in magnitude only during the period of the bipole rotation, suggesting that this rotation is a main contributor to their global variation. Also to be noted, is that the $\alpha$ parameter has the same (negative) sign for both the bipole and full AR.

Summarizing, the evolution of the parameters shown in Figures~\ref{fig:sharp1} and~\ref{fig:sharp2} indicate a clear increase in magnitude, in coincidence with the observed bipole rotation, followed by a decrease that begins as soon as the mini-filament eruptions occur. The similarity between the evolution of the parameters might not be surprising, provided that all of them are computed from spatial derivatives of the observed photospheric field. What is remarkable though, is the clear way in which the timing of the rotation and eruption occurrences mark the steps of the evolution of the parameters, implying a direct relation between the rotation and further energy release and the reconfiguration of the photospheric magnetic field. This suggests that, at least in a case of a rapid field reconfiguration produced by bulk photospheric motions, these parameters are good indicators of soon-to-be-produced flares and eruptions. It is noteworthy that part of the success of these parameters to so clearly reflect the configuration evolution, is due to the fact that they are computed over a selected area that precisely covers the region of interest (the bipole). A computation made over the full AR does not necessarily provide such clear results.  

As we described in Section~\ref{s:helicity}, in Figures~\ref{fig:poynting} and~\ref{fig:helicity} we plot the photospheric injection and accumulated sum of the magnetic energy and magnetic helicity for the full AR (panels a and b) and an area containing only the bipole (panels c and d), respectively. In both figures, the color coding of the curves is blue for the computation of the emergence terms and red for the shear terms, as defined in Section~\ref{s:helicity} (Equations~\ref{eq:energy} and~\ref{eq:helicity}). The black curves correspond to the total injection, which is the sum of those two terms. These injection rates and the corresponding accumulated sums are extensive quantities that produce values that are several orders of magnitude different for the full AR and the bipole. This is the main reason why we plot them in separate panels.

As we did in Figures~\ref{fig:sharp1} and~\ref{fig:sharp2}, in all the panels of Figures~\ref{fig:poynting} and~\ref{fig:helicity} we add vertical dashed lines indicating the interval of the bipole rotation (in yellow) and the times when the eruptions occurred (in green). Despite being smoothed with a 5-point running average the evolution of the injection rates (panels a and c in both figures) is still very noisy. For that reason, except at the times when the rates clearly depart from 0, it is difficult to infer what is the net injection taking place. The panels (b and d) on the right of the figures, corresponding to the accumulated injection integrated in 1 hour intervals as described in Section~\ref{s:helicity}, allow us to have a clearer notion of how the injections evolve. It is worth reminding that the accumulated quantities computed here are related to the actual injection only in an average way, since we compute the rates at a cadence of one per hour, so the cumulative integration is done considering an approximately constant rate between one computation and the next. It is also worth mentioning that the computed injections only indicate the amount of magnetic energy and helicity that should be added to the initial content of these quantities in the AR at the beginning of the analysis, which are of course unknown in this case. Negative energies as observed in some of the panels of Figure~\ref{fig:poynting} are thus not necessarily unrealistic or nonphysical. 

Panels b and d of Figure~\ref{fig:poynting} show a generally sustained energy injection in the full AR and the bipole areas present in both the emergence and shear components, with the exception of a short decreasing interval between $t \approx 30$ and 40 h in the cumulative energy injection of the bipole, followed by a strong injection that coincides with the fastest rotation period of approximately 30 h represented in the panels of Figure~\ref{fig:bipole} and in Figure~\ref{fig:rotation} (between $t \approx $ 45 and 75 h).

The plots of Figure~\ref{fig:helicity} show a qualitatively similar evolution for the magnetic helicity injection in the whole AR (panels a and b) and in the bipole area (panels c and d), with minor differences in details. The panels indicate a net positive helicity injection (black curves) produced by the sum of a larger positive shear component (red curves) and a smaller negative emergence component (blue curves) of lesser magnitude. Both contributions in both cumulative injection curves (panels b and d) show a sustained growing tendency up to the time when the eruptive events occur (green dashed vertical lines). While the negative emergence term continues to grow in magnitude, the shear component injection stops to grow eventually reverting the sign of the injection rate after the events (see how the fluctuating red curves turn negative after the events and, moreover, after the bipole rotation ceases). This change of sign in the shear component of the helicity injection rate, added to the negative emergence component, produces a dramatic change in the total injection as reflected in all the black curves, the ones corresponding to the injection rate (panels a and c) and their respective cumulative values (panels b and d).

Notice that the positive injection of shear helicity is already present before the identified start of the bipole rotation (first vertical dashed yellow line in Figure~\ref{fig:helicity}). This initial injection could be due to the rotation of small magnetic features associated with the initial emergence of the top of the magnetic structure that will form the observed bipole, which is not necessarily reflected in the computations of Figure~\ref{fig:rotation}.

It is worth noting that, since all the computations involving the full AR include the bipole area, it is expected that the evolution on the bipole have some influence over the full AR measurements, being somehow reflected on its evolution. However, the full AR curves are more than one order of magnitude larger than the curves computed for the bipole area. It is hard to say what part of the observed behavior is directly due to the bipole evolution, as it is clear that the contribution of other parts of the AR, which are outside our analysis, must have a non-negligible influence on this evolution. One would expect the effect of the bipole evolution to be more diluted in the evolution observed in the full AR. It is interesting to see though that the evolution of the full AR qualitatively resembles the evolution in the bipole area.
 

\section{Discussion and Conclusions}
\label{s:conclusion}

The results presented in Section~\ref{s:results} provide a consistent scenario for the evolution of the analyzed parameters and their relation to the observed bipole rotation and production of eruptive events. The presence of a rotating bipole in the middle of the AR produces a localized injection of magnetic energy and helicity in the configuration, which reflects in the increase of photospheric parameters that provide a measure of strong field gradients, torsion, and helicity and energy density. This is confirmed by direct measurements of injected energy and helicity showing a similar evolution. The accumulation of magnetic energy in the strongly sheared structure in and around the PIL of the bipole, accompanied by flux cancellation and the formation of a mini-filament, ends up in an instability that ejects the mini-filament material. This material is thus injected by reconnection into the overlying arcade of larger scale AR loops producing the surge eruption. This same process repeats three times along the approximately 16 h of maximum injection of magnetic energy and helicity. The process stops as soon as the large concentration of shear around the PIL and the bipole rotation cease.

\begin{figure} 
\centerline{\includegraphics[width=0.7\textwidth,clip=]{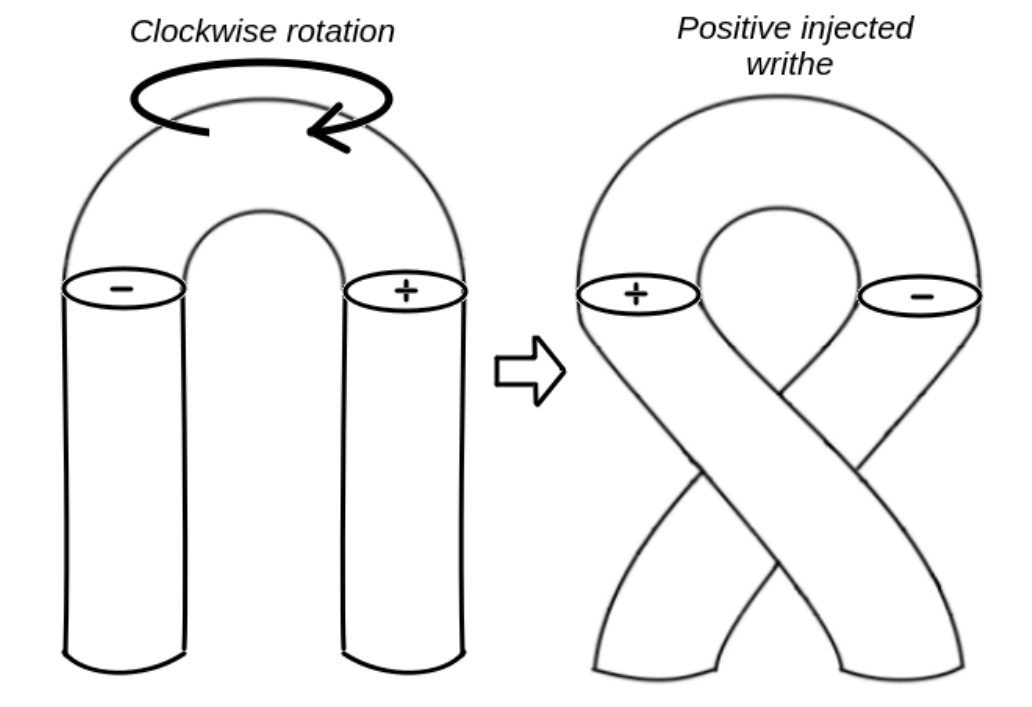}}
\caption{Cartoon showing the deformation suffered by an $\Omega$-loop flux tube (left) whose top is rotated in the clockwise direction (see thick black curved arrow on top of the loop). The drawing on the right shows the resulting deformed flux tube with positive writhe magnetic helicity.}
   \label{fig:pos-writhe}
\end{figure}

Let us now analyze what could possibly be the origin of the observed bipole rotation. There are two possible scenarios to explain this evolution. In the first one, a regular magnetic flux tube in the form of a typical $\Omega$-loop emerges in the middle of the AR with a positive leading polarity and a trailing negative polarity. This is illustrated in the left panel of Figure~\ref{fig:pos-writhe}. If a photospheric vortical motion in the clockwise direction is applied over this bipolar structure, as indicated by the circular thick black arrow on top of the loop, that would produce a deformation of the main axis of the flux tube so that it acquires positive helicity in the form of writhe. Writhe is defined precisely as the magnetic helicity component related to the deformation of the main axis of an $\Omega$-loop flux tube \citep[see e. g.,][]{pevtsov2014}. This clockwise motion is  precisely how the bipole is observed to rotate, and it is also why the shear term of the magnetic helicity injection (see Figure~\ref{fig:helicity}, panels c and d) on the bipole has a positive sign. The rotational motion, as observed in the evolution of the magnetograms, is interpreted by the computational procedure as a positive helicity injection. It is perhaps worth mentioning that the illustrations of Figure~\ref{fig:pos-writhe} are presented as a motivation to interpret the injected helicity sign and they do not pretend to represent the dimensions and proportions of the actual structure.

\begin{figure} 
\centerline{\includegraphics[width=0.7\textwidth,clip=]{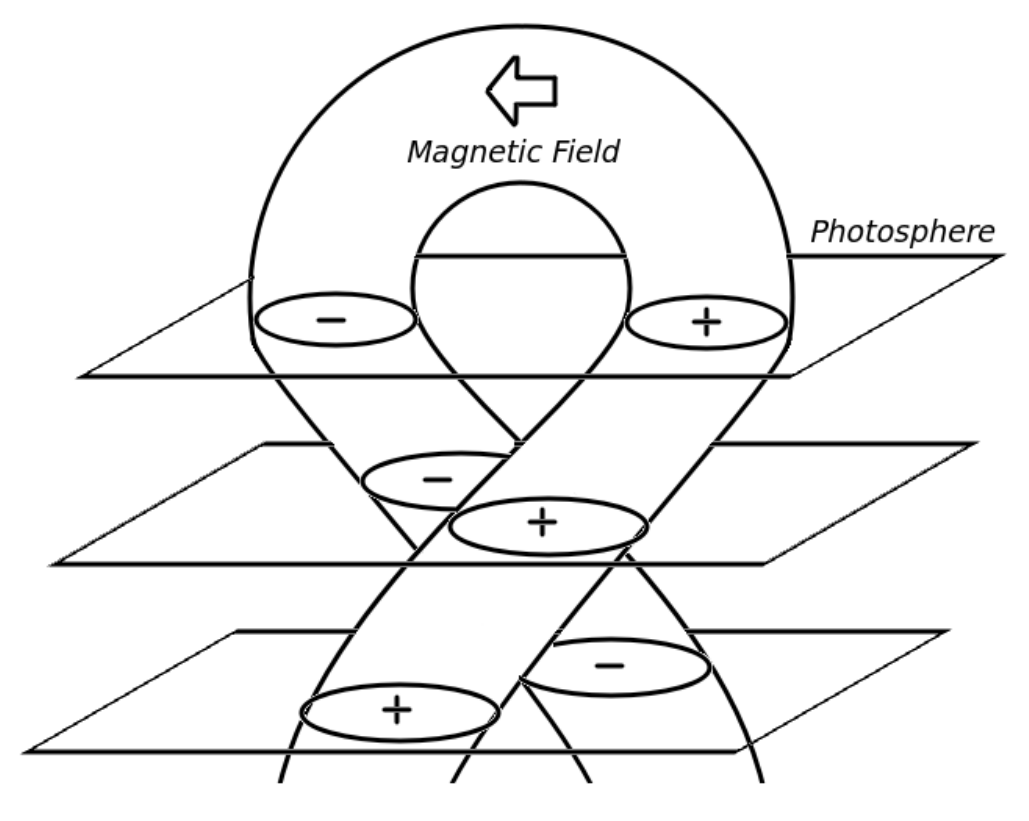}}
\caption{Cartoon representing the progressive emergence of a deformed flux tube with negative writhe helicity. The successive planes show the relative positions of the photosphere as the flux tube emerges. The relative positions of the magnetic polarities where the flux tube cuts the photosphere represent the observed rotation of the bipole that emerged in the middle of the main positive polarity of AR 11476.}
   \label{fig:neg-writhe}
\end{figure}

A second possible scenario is shown in Figure~\ref{fig:neg-writhe}, in which an already deformed flux tube with negative writhe is observed to emerge. As the flux tube emerges the relative positions of the magnetic polarities on the photosphere produce an apparent clockwise rotation coherently with the observations. We emphasize that the mathematical computation given by Equations~\ref{eq:helicity} and successive ones are not able to differentiate between the above two scenarios and it will identify a clockwise rotation with a positive magnetic helicity injection. Once again, the cartoon is for illustrative purposes and does not pretend to be a real representation of the system actual dimensions and proportions. As we discuss in what follows, we consider that the interpretation based on the emergence of a flux tube with negative writhe is more consistent with the overall context.

First of all, AR 11476 is a north hemisphere AR, so according to the Hale-Nicholson law \citep{hale1925} for Solar Cycle 24, its global bipolar magnetic structure is expected to have a negative preceding polarity and a positive following polarity, as it actually occurs, consistently with a global toroidal field orientation from east to west. Regarding the bipole, the magnetic orientation of the feet of the $\Omega$-loop shown in Figure~\ref{fig:pos-writhe} is opposite to the one expected in the north hemisphere for Cycle 24. On the other hand, in the case of the magnetic flux tube with negative writhe shown in Figure~\ref{fig:neg-writhe}, although the apparent orientation at the beginning of the emergence seems to be non-Hale, the ultimate orientation of the feet of the flux tube is consistent with the expected Hale-Nicholson orientation.

Secondly, according to the hemispherical rule of the magnetic helicity sign, northern hemisphere ARs tend to have negative magnetic helicity. This is precisely the case of AR 11476, both globally and for the bipole in particular, as shown in Figure~\ref{fig:sharp2} with the evolution of the force-free $\alpha$ parameter, which is a proxy of the twist helicity. Also, the emergence component of the magnetic helicity injection presented in Figure~\ref{fig:helicity} is consistently negative, supporting an overall negative helicity for the AR and the bipole. As we already discussed, the apparently positive shear injection component can be easily explained by the emergence of a negative writhe flux tube. In support of an overall negative magnetic helicity, we should also mention that the analysis of the topology of the coronal magnetic structure of the whole AR and the region around the bipole, studied in our previous work \citep{lopezfuentes2018}, indicated the systematic presence of negative magnetic twist.

Although the presence of negative magnetic twist in the bipole does not necessarily mean the emergence of a negative writhe flux tube, there exists a mechanism that supports our assumption and can link both facts. The process known as kink instability occurs when a deformed magnetic structure, such as a twisted magnetic flux tube, reaches a threshold above which the magnetic helicity is transferred from twist to writhe, deforming the whole tube structure \citep[see e. g.,][]{fan2021}. Although other mechanisms could be responsible for the flux tube deformation, such as rotational motions in the solar interior before the structure emergence \citep[see e. g.,][]{lopezfuentes2003}, we think that the above arguments support the plausibility of a scenario based on the development of a kink instability. 

Finally, against the possibility of a photospheric vortical motion able to produce the observed clockwise rotation of the bipole, we are not aware of any reported observations of motions of that spatial scale ($\approx$10 Mm) and rotational velocities ($\approx$3 deg h$^{-1}$) sustained over such periods of time (tens of hours). We, therefore, conclude that the scenario described in Figure~\ref{fig:neg-writhe}, based on the emergence of a deformed flux tube, offers an explanation of the observed evolution that is overall more consistent with other known facts.

In this article, we study the evolution of AR 11476 based on the analysis of a series of magnetic parameters and the computation of the injected magnetic energy and helicity. To be able to obtain a complete description of the evolution we need to focus the analysis in a particular sub-region of the AR, namely, a magnetic bipole at whose PIL mini-filament structures were reformed and ejected along a time lapse of approximately 16 h. We confirm that the rotation of this bipolar substructure along several tens of hours was the main mechanism for the injection of magnetic energy and helicity that eventually provided a way for the occurrence of flares and confined eruptions. Our analysis also confirms the conclusion reached by other authors that for a full diagnostic of the flare and CME productivity of an AR it may be necessary to use different proxies for its magnetic evolution \citep{thalmann2021, liokati2022}. Moreover, focusing the analysis in particular sub-regions might be sometimes necessary to fully understand the involved processes \citep[see, e. g.,][and references therein]{ji2023}. These are possibly the main reasons why no single magnetic parameter or measurement has been able to date to provide a definitive proxy for the prediction of active events. The future application of artificial intelligence and machine learning techniques should include ways to extract particular information from localized areas of ARs. The computation of global parameters could be insufficient to provide full diagnostics for the prediction of eruptive events.

In recent years, there has also been a strong effort to find thresholds in some of the analyzed parameters, including injected magnetic helicity, that could be used to discriminate between eruptive and non-eruptive flares \citep[see, e. g.,][]{liu2023}, i. e., flares associated and non-associated to mass ejections into the interplanetary space. Here, we study a case that produced flares associated to confined eruptive phenomena. Although this kind of cases might not appear at first sight too relevant for space weather, the physics involved in the energy injection, instability development, and eruption in a more localized and smaller scale system might provide important clues to bridge the transit from smaller, confined events, to full AR scale phenomena producing the most energetic CMEs.

%
\begin{acks} 
The authors wish to sincerely thank the anonymouos reviewer for useful comments and suggestions that contributed to improve the article. The authors are members of the Carrera del Investigador Cient\'{\i}fico of the Consejo Nacional de Investigaciones Cient\'{\i}ficas y T\'ecnicas (CONICET) of Argentina. The authors acknowledge the use of data from the SDO (NASA) mission. The authors acknowledge Andre Chicrala for sharing his adapted Python version of Peter Schuck's original code DAVE4VM. 
\end{acks}

\begin{authorcontribution}
M.L.F and M.P. contributed to the computation of the magnetic helicity and energy injection. M.L.F. and C.H.M. contributed to the writing of article. All authors reviewed the manuscript.
\end{authorcontribution}
\begin{fundinginformation}
The authors acknowledge financial support from the Argentinian grants PICT 2020-03214 (ANPCyT) and PIP 11220200100985 (CONICET).
\end{fundinginformation}
\begin{dataavailability}
The data used in this work are available from: \\
\url{http://jsoc.stanford.edu/doc/data/hmi/sharp/sharp.htm}.
\end{dataavailability}
%

 \begin{conflict}
The authors declare that they have no conflicts of interest.
 \end{conflict}

%

 \bibliographystyle{spr-mp-sola}
 \bibliography{paper_ar_evol.bib}  

\begin{thebibliography}{66}
\ifx\bisbn     \undefined \def\bisbn  #1{ISBN #1}\fi
\ifx\binits    \undefined \def\binits#1{#1}\fi
\ifx\bauthor   \undefined \def\bauthor#1{#1}\fi
\ifx\batitle   \undefined \def\batitle#1{#1}\fi
\ifx\bjtitle   \undefined \def\bjtitle#1{\textit{#1}}\fi
\ifx\bvolume   \undefined \def\bvolume#1{\textbf{#1}}\fi
\ifx\byear     \undefined \def\byear#1{#1}\fi
\ifx\bissue    \undefined \def\bissue#1{#1}\fi
\ifx\bfpage    \undefined \def\bfpage#1{#1}\fi
\ifx\blpage    \undefined \def\blpage #1{#1}\fi
\ifx\burl      \undefined \def\burl#1{#1}\fi
\ifx\href      \undefined \def\href#1#2{#2}\fi
\ifx\betal     \undefined \def\betal{et al.}\fi
\ifx\bctitle   \undefined \def\bctitle#1{#1}\fi
\ifx\beditor   \undefined \def\beditor#1{#1}\fi
\ifx\bbtitle   \undefined \def\bbtitle#1{\textit{#1}}\fi
\ifx\bedition  \undefined \def\bedition#1{#1}\fi
\ifx\bseriesno \undefined \def\bseriesno#1{\textbf{#1}}\fi
\ifx\blocation \undefined \def\blocation#1{#1}\fi
\ifx\bsertitle \undefined \def\bsertitle#1{\textit{#1}}\fi
\ifx\bsnm      \undefined \def\bsnm#1{#1}\fi
\ifx\bsuffix   \undefined \def\bsuffix#1{#1}\fi
\ifx\bparticle \undefined \def\bparticle#1{#1}\fi
\ifx\barticle  \undefined \def\barticle#1{}\fi
\ifx\binstitute  \undefined \def\binstitute#1{#1}\fi
\ifx\bpublisher  \undefined \def\bpublisher#1{#1}\fi
\ifx\doiurl    \undefined \def\doiurl#1{\href{#1}{DOI}}\fi
\makeatletter
\def\safeHref#1#2#3{\in@{http}{#2}\ifin@\href{#2}{#3}\else\href{#1#2}{#3}\fi}
\makeatother
\ifx\adsurl    \undefined
  \def\adsurl#1{\safeHref{https://ui.adsabs.harvard.edu/abs/}{#1}{ADS}}\fi
\ifx\arxivurl  \undefined
  \def\arxivurl#1{\safeHref{http://arxiv.org/abs/}{#1}{arXiv}}\fi
\ifx\botherref \undefined \def\botherref#1{}\fi
\ifx\url       \undefined \def\url#1{#1}\fi
\ifx\bchapter  \undefined \def\bchapter#1{}\fi
\ifx\bbook     \undefined \def\bbook#1{}\fi
\ifx\bcomment  \undefined \def\bcomment#1{#1}\fi
\ifx\oauthor   \undefined \def\oauthor#1{#1}\fi
\ifx\citeauthoryear \undefined\def \citeauthoryear#1{#1}\fi
\def\endbibitem {}
\ifx\bconflocation  \undefined \def\bconflocation#1{#1} \fi

\bibitem[\protect\citeauthoryear{{Berger}}{1984}]{berger1984}
\begin{barticle}
\bauthor{\bsnm{{Berger}}, \binits{M.A.}}:
\byear{1984},
\batitle{{Rigorous new limits on magnetic helicity dissipation in the solar
  corona}}.
\bjtitle{Geophysical and Astrophysical Fluid Dynamics}
\bvolume{30},
\bfpage{79}.
\doiurl{https://doi.org/10.1080/03091928408210078}.
\adsurl{1984GApFD..30...79B}.
\end{barticle}
\endbibitem

\bibitem[\protect\citeauthoryear{{Berger}}{2003}]{berger2003}
\begin{bchapter}
\bauthor{\bsnm{{Berger}}, \binits{M.A.}}:
\byear{2003},
\bctitle{{Topological quantities in magnetohydrodynamics}}.
In: \beditor{\bsnm{{Ferriz-Mas}}, \binits{A.}},
\beditor{\bsnm{{N{\'u}{\~n}ez}}, \binits{M.}} (eds.)
\bbtitle{Advances in Nonlinear Dynamics},
\bfpage{345}.
\doiurl{https://doi.org/10.1201/9780203493137.ch10}.
\adsurl{2003and..book..345B}.
\end{bchapter}
\endbibitem

\bibitem[\protect\citeauthoryear{{Bobra} and {Couvidat}}{2015}]{bobra2015}
\begin{barticle}
\bauthor{\bsnm{{Bobra}}, \binits{M.G.}},
\bauthor{\bsnm{{Couvidat}}, \binits{S.}}:
\byear{2015},
\batitle{{Solar Flare Prediction Using SDO/HMI Vector Magnetic Field Data with
  a Machine-learning Algorithm}}.
\bjtitle{\apj}
\bvolume{798},
\bfpage{135}.
\doiurl{https://doi.org/10.1088/0004-637X/798/2/135}.
\adsurl{2015ApJ...798..135B}.
\end{barticle}
\endbibitem

\bibitem[\protect\citeauthoryear{{Bobra} and {Ilonidis}}{2016}]{bobra2016}
\begin{barticle}
\bauthor{\bsnm{{Bobra}}, \binits{M.G.}},
\bauthor{\bsnm{{Ilonidis}}, \binits{S.}}:
\byear{2016},
\batitle{{Predicting Coronal Mass Ejections Using Machine Learning Methods}}.
\bjtitle{\apj}
\bvolume{821},
\bfpage{127}.
\doiurl{https://doi.org/10.3847/0004-637X/821/2/127}.
\adsurl{2016ApJ...821..127B}.
\end{barticle}
\endbibitem

\bibitem[\protect\citeauthoryear{{Bobra} et~al.}{2014}]{bobra2014}
\begin{barticle}
\bauthor{\bsnm{{Bobra}}, \binits{M.G.}},
\bauthor{\bsnm{{Sun}}, \binits{X.}},
\bauthor{\bsnm{{Hoeksema}}, \binits{J.T.}},
\bauthor{\bsnm{{Turmon}}, \binits{M.}},
\bauthor{\bsnm{{Liu}}, \binits{Y.}},
\bauthor{\bsnm{{Hayashi}}, \binits{K.}},
\bauthor{\bsnm{{Barnes}}, \binits{G.}},
\bauthor{\bsnm{{Leka}}, \binits{K.D.}}:
\byear{2014},
\batitle{{The Helioseismic and Magnetic Imager (HMI) Vector Magnetic Field
  Pipeline: SHARPs - Space-Weather HMI Active Region Patches}}.
\bjtitle{\solphys}
\bvolume{289},
\bfpage{3549}.
\doiurl{https://doi.org/10.1007/s11207-014-0529-3}.
\adsurl{2014SoPh..289.3549B}.
\end{barticle}
\endbibitem

\bibitem[\protect\citeauthoryear{{Bobra} et~al.}{2021}]{bobra2021}
\begin{barticle}
\bauthor{\bsnm{{Bobra}}, \binits{M.G.}},
\bauthor{\bsnm{{Wright}}, \binits{P.J.}},
\bauthor{\bsnm{{Sun}}, \binits{X.}},
\bauthor{\bsnm{{Turmon}}, \binits{M.J.}}:
\byear{2021},
\batitle{{SMARPs and SHARPs: Two Solar Cycles of Active Region Data}}.
\bjtitle{\apjs}
\bvolume{256},
\bfpage{26}.
\doiurl{https://doi.org/10.3847/1538-4365/ac1f1d}.
\adsurl{2021ApJS..256...26B}.
\end{barticle}
\endbibitem

\bibitem[\protect\citeauthoryear{{D{\'e}moulin} and
  {Pariat}}{2009}]{demoulin2009}
\begin{barticle}
\bauthor{\bsnm{{D{\'e}moulin}}, \binits{P.}},
\bauthor{\bsnm{{Pariat}}, \binits{E.}}:
\byear{2009},
\batitle{{Modelling and observations of photospheric magnetic helicity}}.
\bjtitle{Advances in Space Research}
\bvolume{43},
\bfpage{1013}.
\doiurl{https://doi.org/10.1016/j.asr.2008.12.004}.
\adsurl{2009AdSpR..43.1013D}.
\end{barticle}
\endbibitem

\bibitem[\protect\citeauthoryear{{Fan}}{2021}]{fan2021}
\begin{barticle}
\bauthor{\bsnm{{Fan}}, \binits{Y.}}:
\byear{2021},
\batitle{{Magnetic fields in the solar convection zone}}.
\bjtitle{Living Reviews in Solar Physics}
\bvolume{18},
\bfpage{5}.
\doiurl{https://doi.org/10.1007/s41116-021-00031-2}.
\adsurl{2021LRSP...18....5F}.
\end{barticle}
\endbibitem

\bibitem[\protect\citeauthoryear{{Georgoulis} et~al.}{2021}]{georgoulis2021}
\begin{barticle}
\bauthor{\bsnm{{Georgoulis}}, \binits{M.K.}},
\bauthor{\bsnm{{Bloomfield}}, \binits{D.S.}},
\bauthor{\bsnm{{Piana}}, \binits{M.}},
\bauthor{\bsnm{{Massone}}, \binits{A.M.}},
\bauthor{\bsnm{{Soldati}}, \binits{M.}},
\bauthor{\bsnm{{Gallagher}}, \binits{P.T.}},
\bauthor{\bsnm{{Pariat}}, \binits{E.}},
\bauthor{\bsnm{{Vilmer}}, \binits{N.}},
\bauthor{\bsnm{{Buchlin}}, \binits{E.}},
\bauthor{\bsnm{{Baudin}}, \binits{F.}},
\bauthor{\bsnm{{Csillaghy}}, \binits{A.}},
\bauthor{\bsnm{{Sathiapal}}, \binits{H.}},
\bauthor{\bsnm{{Jackson}}, \binits{D.R.}},
\bauthor{\bsnm{{Alingery}}, \binits{P.}},
\bauthor{\bsnm{{Benvenuto}}, \binits{F.}},
\bauthor{\bsnm{{Campi}}, \binits{C.}},
\bauthor{\bsnm{{Florios}}, \binits{K.}},
\bauthor{\bsnm{{Gontikakis}}, \binits{C.}},
\bauthor{\bsnm{{Guennou}}, \binits{C.}},
\bauthor{\bsnm{{Guerra}}, \binits{J.A.}},
\bauthor{\bsnm{{Kontogiannis}}, \binits{I.}},
\bauthor{\bsnm{{Latorre}}, \binits{V.}},
\bauthor{\bsnm{{Murray}}, \binits{S.A.}},
\bauthor{\bsnm{{Park}}, \binits{S.-H.}},
\bauthor{\bsnm{{von Stachelski}}, \binits{S.}},
\bauthor{\bsnm{{Torbica}}, \binits{A.}},
\bauthor{\bsnm{{Vischi}}, \binits{D.}},
\bauthor{\bsnm{{Worsfold}}, \binits{M.}}:
\byear{2021},
\batitle{{The flare likelihood and region eruption forecasting (FLARECAST)
  project: flare forecasting in the big data \& machine learning era}}.
\bjtitle{Journal of Space Weather and Space Climate}
\bvolume{11},
\bfpage{39}.
\doiurl{https://doi.org/10.1051/swsc/2021023}.
\adsurl{2021JSWSC..11...39G}.
\end{barticle}
\endbibitem

\bibitem[\protect\citeauthoryear{{Green}, {Kliem}, and
  {Wallace}}{2011}]{green2011}
\begin{barticle}
\bauthor{\bsnm{{Green}}, \binits{L.M.}},
\bauthor{\bsnm{{Kliem}}, \binits{B.}},
\bauthor{\bsnm{{Wallace}}, \binits{A.J.}}:
\byear{2011},
\batitle{{Photospheric flux cancellation and associated flux rope formation and
  eruption}}.
\bjtitle{\aap}
\bvolume{526},
\bfpage{A2}.
\doiurl{https://doi.org/10.1051/0004-6361/201015146}.
\adsurl{2011A&A...526A...2G}.
\end{barticle}
\endbibitem

\bibitem[\protect\citeauthoryear{{Green} et~al.}{2002}]{green2002}
\begin{barticle}
\bauthor{\bsnm{{Green}}, \binits{L.M.}},
\bauthor{\bsnm{{L{\'o}pez fuentes}}, \binits{M.C.}},
\bauthor{\bsnm{{Mandrini}}, \binits{C.H.}},
\bauthor{\bsnm{{D{\'e}moulin}}, \binits{P.}},
\bauthor{\bsnm{{Van Driel-Gesztelyi}}, \binits{L.}},
\bauthor{\bsnm{{Culhane}}, \binits{J.L.}}:
\byear{2002},
\batitle{{The Magnetic Helicity Budget of a cme-Prolific Active Region}}.
\bjtitle{\solphys}
\bvolume{208},
\bfpage{43}.
\doiurl{https://doi.org/10.1023/A:1019658520033}.
\adsurl{2002SoPh..208...43G}.
\end{barticle}
\endbibitem

\bibitem[\protect\citeauthoryear{{Green} et~al.}{2018}]{green2018}
\begin{barticle}
\bauthor{\bsnm{{Green}}, \binits{L.M.}},
\bauthor{\bsnm{{T{\"o}r{\"o}k}}, \binits{T.}},
\bauthor{\bsnm{{Vr{\v{s}}nak}}, \binits{B.}},
\bauthor{\bsnm{{Manchester}}, \binits{W.}},
\bauthor{\bsnm{{Veronig}}, \binits{A.}}:
\byear{2018},
\batitle{{The Origin, Early Evolution and Predictability of Solar Eruptions}}.
\bjtitle{\ssr}
\bvolume{214},
\bfpage{46}.
\doiurl{https://doi.org/10.1007/s11214-017-0462-5}.
\adsurl{2018SSRv..214...46G}.
\end{barticle}
\endbibitem

\bibitem[\protect\citeauthoryear{{Gupta}, {Thalmann}, and
  {Veronig}}{2021}]{gupta2021}
\begin{barticle}
\bauthor{\bsnm{{Gupta}}, \binits{M.}},
\bauthor{\bsnm{{Thalmann}}, \binits{J.K.}},
\bauthor{\bsnm{{Veronig}}, \binits{A.M.}}:
\byear{2021},
\batitle{{Magnetic helicity and energy budget around large confined and
  eruptive solar flares}}.
\bjtitle{\aap}
\bvolume{653},
\bfpage{A69}.
\doiurl{https://doi.org/10.1051/0004-6361/202140591}.
\adsurl{2021A&A...653A..69G}.
\end{barticle}
\endbibitem

\bibitem[\protect\citeauthoryear{{Hagyard} et~al.}{1984}]{hagyard1984}
\begin{barticle}
\bauthor{\bsnm{{Hagyard}}, \binits{M.J.}},
\bauthor{\bsnm{{Smith}}, \binits{J.} \bsuffix{J.~B.}},
\bauthor{\bsnm{{Teuber}}, \binits{D.}},
\bauthor{\bsnm{{West}}, \binits{E.A.}}:
\byear{1984},
\batitle{{A Quantitative Study Relating Observed Shear in Photospheric Magnetic
  Fields to Repeated Flaring}}.
\bjtitle{\solphys}
\bvolume{91},
\bfpage{115}.
\doiurl{https://doi.org/10.1007/BF00213618}.
\adsurl{1984SoPh...91..115H}.
\end{barticle}
\endbibitem

\bibitem[\protect\citeauthoryear{{Hale} and {Nicholson}}{1925}]{hale1925}
\begin{barticle}
\bauthor{\bsnm{{Hale}}, \binits{G.E.}},
\bauthor{\bsnm{{Nicholson}}, \binits{S.B.}}:
\byear{1925},
\batitle{{The Law of Sun-Spot Polarity}}.
\bjtitle{\apj}
\bvolume{62},
\bfpage{270}.
\doiurl{https://doi.org/10.1086/142933}.
\adsurl{1925ApJ....62..270H}.
\end{barticle}
\endbibitem

\bibitem[\protect\citeauthoryear{{Ji} et~al.}{2023}]{ji2023}
\begin{barticle}
\bauthor{\bsnm{{Ji}}, \binits{A.}},
\bauthor{\bsnm{{Cai}}, \binits{X.}},
\bauthor{\bsnm{{Khasayeva}}, \binits{N.}},
\bauthor{\bsnm{{Georgoulis}}, \binits{M.K.}},
\bauthor{\bsnm{{Martens}}, \binits{P.C.}},
\bauthor{\bsnm{{Angryk}}, \binits{R.A.}},
\bauthor{\bsnm{{Aydin}}, \binits{B.}}:
\byear{2023},
\batitle{{A Systematic Magnetic Polarity Inversion Line Data Set from SDO/HMI
  Magnetograms}}.
\bjtitle{\apjs}
\bvolume{265},
\bfpage{28}.
\doiurl{https://doi.org/10.3847/1538-4365/acb43a}.
\adsurl{2023ApJS..265...28J}.
\end{barticle}
\endbibitem

\bibitem[\protect\citeauthoryear{{Kontogiannis}}{2023}]{kontogiannis2023}
\begin{barticle}
\bauthor{\bsnm{{Kontogiannis}}, \binits{I.}}:
\byear{2023},
\batitle{{The characteristics of flare- and CME-productive solar active
  regions}}.
\bjtitle{Advances in Space Research}
\bvolume{71},
\bfpage{2017}.
\doiurl{https://doi.org/10.1016/j.asr.2022.10.008}.
\adsurl{2023AdSpR..71.2017K}.
\end{barticle}
\endbibitem

\bibitem[\protect\citeauthoryear{{Kusano} et~al.}{2002}]{kusano2002}
\begin{barticle}
\bauthor{\bsnm{{Kusano}}, \binits{K.}},
\bauthor{\bsnm{{Maeshiro}}, \binits{T.}},
\bauthor{\bsnm{{Yokoyama}}, \binits{T.}},
\bauthor{\bsnm{{Sakurai}}, \binits{T.}}:
\byear{2002},
\batitle{{Measurement of Magnetic Helicity Injection and Free Energy Loading
  into the Solar Corona}}.
\bjtitle{\apj}
\bvolume{577},
\bfpage{501}.
\doiurl{https://doi.org/10.1086/342171}.
\adsurl{2002ApJ...577..501K}.
\end{barticle}
\endbibitem

\bibitem[\protect\citeauthoryear{{LaBonte}, {Georgoulis}, and
  {Rust}}{2007}]{labonte2007}
\begin{barticle}
\bauthor{\bsnm{{LaBonte}}, \binits{B.J.}},
\bauthor{\bsnm{{Georgoulis}}, \binits{M.K.}},
\bauthor{\bsnm{{Rust}}, \binits{D.M.}}:
\byear{2007},
\batitle{{Survey of Magnetic Helicity Injection in Regions Producing X-Class
  Flares}}.
\bjtitle{\apj}
\bvolume{671},
\bfpage{955}.
\doiurl{https://doi.org/10.1086/522682}.
\adsurl{2007ApJ...671..955L}.
\end{barticle}
\endbibitem

\bibitem[\protect\citeauthoryear{{Leka} and {Barnes}}{2007}]{leka2007}
\begin{barticle}
\bauthor{\bsnm{{Leka}}, \binits{K.D.}},
\bauthor{\bsnm{{Barnes}}, \binits{G.}}:
\byear{2007},
\batitle{{Photospheric Magnetic Field Properties of Flaring versus Flare-quiet
  Active Regions. IV. A Statistically Significant Sample}}.
\bjtitle{\apj}
\bvolume{656},
\bfpage{1173}.
\doiurl{https://doi.org/10.1086/510282}.
\adsurl{2007ApJ...656.1173L}.
\end{barticle}
\endbibitem

\bibitem[\protect\citeauthoryear{{Leka} et~al.}{2019a}]{leka2019a}
\begin{barticle}
\bauthor{\bsnm{{Leka}}, \binits{K.D.}},
\bauthor{\bsnm{{Park}}, \binits{S.-H.}},
\bauthor{\bsnm{{Kusano}}, \binits{K.}},
\bauthor{\bsnm{{Andries}}, \binits{J.}},
\bauthor{\bsnm{{Barnes}}, \binits{G.}},
\bauthor{\bsnm{{Bingham}}, \binits{S.}},
\bauthor{\bsnm{{Bloomfield}}, \binits{D.S.}},
\bauthor{\bsnm{{McCloskey}}, \binits{A.E.}},
\bauthor{\bsnm{{Delouille}}, \binits{V.}},
\bauthor{\bsnm{{Falconer}}, \binits{D.}},
\bauthor{\bsnm{{Gallagher}}, \binits{P.T.}},
\bauthor{\bsnm{{Georgoulis}}, \binits{M.K.}},
\bauthor{\bsnm{{Kubo}}, \binits{Y.}},
\bauthor{\bsnm{{Lee}}, \binits{K.}},
\bauthor{\bsnm{{Lee}}, \binits{S.}},
\bauthor{\bsnm{{Lobzin}}, \binits{V.}},
\bauthor{\bsnm{{Mun}}, \binits{J.}},
\bauthor{\bsnm{{Murray}}, \binits{S.A.}},
\bauthor{\bsnm{{Hamad Nageem}}, \binits{T.A.M.}},
\bauthor{\bsnm{{Qahwaji}}, \binits{R.}},
\bauthor{\bsnm{{Sharpe}}, \binits{M.}},
\bauthor{\bsnm{{Steenburgh}}, \binits{R.A.}},
\bauthor{\bsnm{{Steward}}, \binits{G.}},
\bauthor{\bsnm{{Terkildsen}}, \binits{M.}}:
\byear{2019}a,
\batitle{{A Comparison of Flare Forecasting Methods. II. Benchmarks, Metrics,
  and Performance Results for Operational Solar Flare Forecasting Systems}}.
\bjtitle{\apjs}
\bvolume{243},
\bfpage{36}.
\doiurl{https://doi.org/10.3847/1538-4365/ab2e12}.
\adsurl{2019ApJS..243...36L}.
\end{barticle}
\endbibitem

\bibitem[\protect\citeauthoryear{{Leka} et~al.}{2019b}]{leka2019b}
\begin{barticle}
\bauthor{\bsnm{{Leka}}, \binits{K.D.}},
\bauthor{\bsnm{{Park}}, \binits{S.-H.}},
\bauthor{\bsnm{{Kusano}}, \binits{K.}},
\bauthor{\bsnm{{Andries}}, \binits{J.}},
\bauthor{\bsnm{{Barnes}}, \binits{G.}},
\bauthor{\bsnm{{Bingham}}, \binits{S.}},
\bauthor{\bsnm{{Bloomfield}}, \binits{D.S.}},
\bauthor{\bsnm{{McCloskey}}, \binits{A.E.}},
\bauthor{\bsnm{{Delouille}}, \binits{V.}},
\bauthor{\bsnm{{Falconer}}, \binits{D.}},
\bauthor{\bsnm{{Gallagher}}, \binits{P.T.}},
\bauthor{\bsnm{{Georgoulis}}, \binits{M.K.}},
\bauthor{\bsnm{{Kubo}}, \binits{Y.}},
\bauthor{\bsnm{{Lee}}, \binits{K.}},
\bauthor{\bsnm{{Lee}}, \binits{S.}},
\bauthor{\bsnm{{Lobzin}}, \binits{V.}},
\bauthor{\bsnm{{Mun}}, \binits{J.}},
\bauthor{\bsnm{{Murray}}, \binits{S.A.}},
\bauthor{\bsnm{{Hamad Nageem}}, \binits{T.A.M.}},
\bauthor{\bsnm{{Qahwaji}}, \binits{R.}},
\bauthor{\bsnm{{Sharpe}}, \binits{M.}},
\bauthor{\bsnm{{Steenburgh}}, \binits{R.A.}},
\bauthor{\bsnm{{Steward}}, \binits{G.}},
\bauthor{\bsnm{{Terkildsen}}, \binits{M.}}:
\byear{2019}b,
\batitle{{A Comparison of Flare Forecasting Methods. III. Systematic Behaviors
  of Operational Solar Flare Forecasting Systems}}.
\bjtitle{\apj}
\bvolume{881},
\bfpage{101}.
\doiurl{https://doi.org/10.3847/1538-4357/ab2e11}.
\adsurl{2019ApJ...881..101L}.
\end{barticle}
\endbibitem

\bibitem[\protect\citeauthoryear{{Lemen} et~al.}{2012}]{lemen2012}
\begin{barticle}
\bauthor{\bsnm{{Lemen}}, \binits{J.R.}},
\bauthor{\bsnm{{Title}}, \binits{A.M.}},
\bauthor{\bsnm{{Akin}}, \binits{D.J.}},
\bauthor{\bsnm{{Boerner}}, \binits{P.F.}},
\bauthor{\bsnm{{Chou}}, \binits{C.}},
\bauthor{\bsnm{{Drake}}, \binits{J.F.}},
\bauthor{\bsnm{{Duncan}}, \binits{D.W.}},
\bauthor{\bsnm{{Edwards}}, \binits{C.G.}},
\bauthor{\bsnm{{Friedlaender}}, \binits{F.M.}},
\bauthor{\bsnm{{Heyman}}, \binits{G.F.}},
\bauthor{\bsnm{{Hurlburt}}, \binits{N.E.}},
\bauthor{\bsnm{{Katz}}, \binits{N.L.}},
\bauthor{\bsnm{{Kushner}}, \binits{G.D.}},
\bauthor{\bsnm{{Levay}}, \binits{M.}},
\bauthor{\bsnm{{Lindgren}}, \binits{R.W.}},
\bauthor{\bsnm{{Mathur}}, \binits{D.P.}},
\bauthor{\bsnm{{McFeaters}}, \binits{E.L.}},
\bauthor{\bsnm{{Mitchell}}, \binits{S.}},
\bauthor{\bsnm{{Rehse}}, \binits{R.A.}},
\bauthor{\bsnm{{Schrijver}}, \binits{C.J.}},
\bauthor{\bsnm{{Springer}}, \binits{L.A.}},
\bauthor{\bsnm{{Stern}}, \binits{R.A.}},
\bauthor{\bsnm{{Tarbell}}, \binits{T.D.}},
\bauthor{\bsnm{{Wuelser}}, \binits{J.-P.}},
\bauthor{\bsnm{{Wolfson}}, \binits{C.J.}},
\bauthor{\bsnm{{Yanari}}, \binits{C.}},
\bauthor{\bsnm{{Bookbinder}}, \binits{J.A.}},
\bauthor{\bsnm{{Cheimets}}, \binits{P.N.}},
\bauthor{\bsnm{{Caldwell}}, \binits{D.}},
\bauthor{\bsnm{{Deluca}}, \binits{E.E.}},
\bauthor{\bsnm{{Gates}}, \binits{R.}},
\bauthor{\bsnm{{Golub}}, \binits{L.}},
\bauthor{\bsnm{{Park}}, \binits{S.}},
\bauthor{\bsnm{{Podgorski}}, \binits{W.A.}},
\bauthor{\bsnm{{Bush}}, \binits{R.I.}},
\bauthor{\bsnm{{Scherrer}}, \binits{P.H.}},
\bauthor{\bsnm{{Gummin}}, \binits{M.A.}},
\bauthor{\bsnm{{Smith}}, \binits{P.}},
\bauthor{\bsnm{{Auker}}, \binits{G.}},
\bauthor{\bsnm{{Jerram}}, \binits{P.}},
\bauthor{\bsnm{{Pool}}, \binits{P.}},
\bauthor{\bsnm{{Soufli}}, \binits{R.}},
\bauthor{\bsnm{{Windt}}, \binits{D.L.}},
\bauthor{\bsnm{{Beardsley}}, \binits{S.}},
\bauthor{\bsnm{{Clapp}}, \binits{M.}},
\bauthor{\bsnm{{Lang}}, \binits{J.}},
\bauthor{\bsnm{{Waltham}}, \binits{N.}}:
\byear{2012},
\batitle{{The Atmospheric Imaging Assembly (AIA) on the Solar Dynamics
  Observatory (SDO)}}.
\bjtitle{\solphys}
\bvolume{275},
\bfpage{17}.
\doiurl{https://doi.org/10.1007/s11207-011-9776-8}.
\adsurl{2012SoPh..275...17L}.
\end{barticle}
\endbibitem

\bibitem[\protect\citeauthoryear{{Liokati}, {Nindos}, and
  {Liu}}{2022}]{liokati2022}
\begin{barticle}
\bauthor{\bsnm{{Liokati}}, \binits{E.}},
\bauthor{\bsnm{{Nindos}}, \binits{A.}},
\bauthor{\bsnm{{Liu}}, \binits{Y.}}:
\byear{2022},
\batitle{{Magnetic helicity and energy of emerging solar active regions and
  their erruptivity}}.
\bjtitle{\aap}
\bvolume{662},
\bfpage{A6}.
\doiurl{https://doi.org/10.1051/0004-6361/202142868}.
\adsurl{2022A&A...662A...6L}.
\end{barticle}
\endbibitem

\bibitem[\protect\citeauthoryear{{Liu} and {Schuck}}{2012}]{liu2012}
\begin{barticle}
\bauthor{\bsnm{{Liu}}, \binits{Y.}},
\bauthor{\bsnm{{Schuck}}, \binits{P.W.}}:
\byear{2012},
\batitle{{Magnetic Energy and Helicity in Two Emerging Active Regions in the
  Sun}}.
\bjtitle{\apj}
\bvolume{761},
\bfpage{105}.
\doiurl{https://doi.org/10.1088/0004-637X/761/2/105}.
\adsurl{2012ApJ...761..105L}.
\end{barticle}
\endbibitem

\bibitem[\protect\citeauthoryear{{Liu} et~al.}{2023}]{liu2023}
\begin{barticle}
\bauthor{\bsnm{{Liu}}, \binits{Y.}},
\bauthor{\bsnm{{Welsch}}, \binits{B.T.}},
\bauthor{\bsnm{{Valori}}, \binits{G.}},
\bauthor{\bsnm{{Georgoulis}}, \binits{M.K.}},
\bauthor{\bsnm{{Guo}}, \binits{Y.}},
\bauthor{\bsnm{{Pariat}}, \binits{E.}},
\bauthor{\bsnm{{Park}}, \binits{S.-H.}},
\bauthor{\bsnm{{Thalmann}}, \binits{J.K.}}:
\byear{2023},
\batitle{{Changes of Magnetic Energy and Helicity in Solar Active Regions from
  Major Flares}}.
\bjtitle{\apj}
\bvolume{942},
\bfpage{27}.
\doiurl{https://doi.org/10.3847/1538-4357/aca3a6}.
\adsurl{2023ApJ...942...27L}.
\end{barticle}
\endbibitem

\bibitem[\protect\citeauthoryear{{L{\'o}pez Fuentes}
  et~al.}{2000}]{lopezfuentes2000}
\begin{barticle}
\bauthor{\bsnm{{L{\'o}pez Fuentes}}, \binits{M.C.}},
\bauthor{\bsnm{{Demoulin}}, \binits{P.}},
\bauthor{\bsnm{{Mandrini}}, \binits{C.H.}},
\bauthor{\bsnm{{van Driel-Gesztelyi}}, \binits{L.}}:
\byear{2000},
\batitle{{The Counterkink Rotation of a Non-Hale Active Region}}.
\bjtitle{\apj}
\bvolume{544},
\bfpage{540}.
\doiurl{https://doi.org/10.1086/317180}.
\adsurl{2000ApJ...544..540L}.
\end{barticle}
\endbibitem

\bibitem[\protect\citeauthoryear{{L{\'o}pez Fuentes}
  et~al.}{2003}]{lopezfuentes2003}
\begin{barticle}
\bauthor{\bsnm{{L{\'o}pez Fuentes}}, \binits{M.C.}},
\bauthor{\bsnm{{D{\'e}moulin}}, \binits{P.}},
\bauthor{\bsnm{{Mandrini}}, \binits{C.H.}},
\bauthor{\bsnm{{Pevtsov}}, \binits{A.A.}},
\bauthor{\bsnm{{van Driel-Gesztelyi}}, \binits{L.}}:
\byear{2003},
\batitle{{Magnetic twist and writhe of active regions. On the origin of
  deformed flux tubes}}.
\bjtitle{\aap}
\bvolume{397},
\bfpage{305}.
\doiurl{https://doi.org/10.1051/0004-6361:20021487}.
\adsurl{2003A&A...397..305L}.
\end{barticle}
\endbibitem

\bibitem[\protect\citeauthoryear{{L{\'o}pez Fuentes}
  et~al.}{2018}]{lopezfuentes2018}
\begin{barticle}
\bauthor{\bsnm{{L{\'o}pez Fuentes}}, \binits{M.}},
\bauthor{\bsnm{{Mandrini}}, \binits{C.H.}},
\bauthor{\bsnm{{Poisson}}, \binits{M.}},
\bauthor{\bsnm{{D{\'e}moulin}}, \binits{P.}},
\bauthor{\bsnm{{Cristiani}}, \binits{G.}},
\bauthor{\bsnm{{L{\'o}pez}}, \binits{F.M.}},
\bauthor{\bsnm{{Luoni}}, \binits{M.L.}}:
\byear{2018},
\batitle{{Physical Processes Involved in the EUV ``Surge'' Event of 9 May
  2012}}.
\bjtitle{\solphys}
\bvolume{293},
\bfpage{166}.
\doiurl{https://doi.org/10.1007/s11207-018-1384-4}.
\adsurl{2018SoPh..293..166L}.
\end{barticle}
\endbibitem

\bibitem[\protect\citeauthoryear{{Low}}{1996}]{low1996}
\begin{barticle}
\bauthor{\bsnm{{Low}}, \binits{B.C.}}:
\byear{1996},
\batitle{{Solar Activity and the Corona}}.
\bjtitle{\solphys}
\bvolume{167},
\bfpage{217}.
\doiurl{https://doi.org/10.1007/BF00146338}.
\adsurl{1996SoPh..167..217L}.
\end{barticle}
\endbibitem

\bibitem[\protect\citeauthoryear{{Mandrini} et~al.}{2005}]{mandrini2005}
\begin{barticle}
\bauthor{\bsnm{{Mandrini}}, \binits{C.H.}},
\bauthor{\bsnm{{Pohjolainen}}, \binits{S.}},
\bauthor{\bsnm{{Dasso}}, \binits{S.}},
\bauthor{\bsnm{{Green}}, \binits{L.M.}},
\bauthor{\bsnm{{D{\'e}moulin}}, \binits{P.}},
\bauthor{\bsnm{{van Driel-Gesztelyi}}, \binits{L.}},
\bauthor{\bsnm{{Copperwheat}}, \binits{C.}},
\bauthor{\bsnm{{Foley}}, \binits{C.}}:
\byear{2005},
\batitle{{Interplanetary flux rope ejected from an X-ray bright point. The
  smallest magnetic cloud source-region ever observed}}.
\bjtitle{\aap}
\bvolume{434},
\bfpage{725}.
\doiurl{https://doi.org/10.1051/0004-6361:20041079}.
\adsurl{2005A&A...434..725M}.
\end{barticle}
\endbibitem

\bibitem[\protect\citeauthoryear{{Martens} and {Zwaan}}{2001}]{martens2001}
\begin{barticle}
\bauthor{\bsnm{{Martens}}, \binits{P.C.}},
\bauthor{\bsnm{{Zwaan}}, \binits{C.}}:
\byear{2001},
\batitle{{Origin and Evolution of Filament-Prominence Systems}}.
\bjtitle{\apj}
\bvolume{558},
\bfpage{872}.
\doiurl{https://doi.org/10.1086/322279}.
\adsurl{2001ApJ...558..872M}.
\end{barticle}
\endbibitem

\bibitem[\protect\citeauthoryear{{Nishizuka} et~al.}{2017}]{nishizuka2017}
\begin{barticle}
\bauthor{\bsnm{{Nishizuka}}, \binits{N.}},
\bauthor{\bsnm{{Sugiura}}, \binits{K.}},
\bauthor{\bsnm{{Kubo}}, \binits{Y.}},
\bauthor{\bsnm{{Den}}, \binits{M.}},
\bauthor{\bsnm{{Watari}}, \binits{S.}},
\bauthor{\bsnm{{Ishii}}, \binits{M.}}:
\byear{2017},
\batitle{{Solar Flare Prediction Model with Three Machine-learning Algorithms
  using Ultraviolet Brightening and Vector Magnetograms}}.
\bjtitle{\apj}
\bvolume{835},
\bfpage{156}.
\doiurl{https://doi.org/10.3847/1538-4357/835/2/156}.
\adsurl{2017ApJ...835..156N}.
\end{barticle}
\endbibitem

\bibitem[\protect\citeauthoryear{{Panesar}, {Sterling}, and
  {Moore}}{2017}]{panesar2017}
\begin{barticle}
\bauthor{\bsnm{{Panesar}}, \binits{N.K.}},
\bauthor{\bsnm{{Sterling}}, \binits{A.C.}},
\bauthor{\bsnm{{Moore}}, \binits{R.L.}}:
\byear{2017},
\batitle{{Magnetic Flux Cancellation as the Origin of Solar Quiet-region
  Pre-jet Minifilaments}}.
\bjtitle{\apj}
\bvolume{844},
\bfpage{131}.
\doiurl{https://doi.org/10.3847/1538-4357/aa7b77}.
\adsurl{2017ApJ...844..131P}.
\end{barticle}
\endbibitem

\bibitem[\protect\citeauthoryear{{Pariat}, {D{\'e}moulin}, and
  {Berger}}{2005}]{pariat2005}
\begin{barticle}
\bauthor{\bsnm{{Pariat}}, \binits{E.}},
\bauthor{\bsnm{{D{\'e}moulin}}, \binits{P.}},
\bauthor{\bsnm{{Berger}}, \binits{M.A.}}:
\byear{2005},
\batitle{{Photospheric flux density of magnetic helicity}}.
\bjtitle{\aap}
\bvolume{439},
\bfpage{1191}.
\doiurl{https://doi.org/10.1051/0004-6361:20052663}.
\adsurl{2005A&A...439.1191P}.
\end{barticle}
\endbibitem

\bibitem[\protect\citeauthoryear{{Pariat} et~al.}{2015}]{pariat2015}
\begin{barticle}
\bauthor{\bsnm{{Pariat}}, \binits{E.}},
\bauthor{\bsnm{{Valori}}, \binits{G.}},
\bauthor{\bsnm{{D{\'e}moulin}}, \binits{P.}},
\bauthor{\bsnm{{Dalmasse}}, \binits{K.}}:
\byear{2015},
\batitle{{Testing magnetic helicity conservation in a solar-like active
  event}}.
\bjtitle{\aap}
\bvolume{580},
\bfpage{A128}.
\doiurl{https://doi.org/10.1051/0004-6361/201525811}.
\adsurl{2015A&A...580A.128P}.
\end{barticle}
\endbibitem

\bibitem[\protect\citeauthoryear{{Pariat} et~al.}{2017}]{pariat2017}
\begin{barticle}
\bauthor{\bsnm{{Pariat}}, \binits{E.}},
\bauthor{\bsnm{{Leake}}, \binits{J.E.}},
\bauthor{\bsnm{{Valori}}, \binits{G.}},
\bauthor{\bsnm{{Linton}}, \binits{M.G.}},
\bauthor{\bsnm{{Zuccarello}}, \binits{F.P.}},
\bauthor{\bsnm{{Dalmasse}}, \binits{K.}}:
\byear{2017},
\batitle{{Relative magnetic helicity as a diagnostic of solar eruptivity}}.
\bjtitle{\aap}
\bvolume{601},
\bfpage{A125}.
\doiurl{https://doi.org/10.1051/0004-6361/201630043}.
\adsurl{2017A&A...601A.125P}.
\end{barticle}
\endbibitem

\bibitem[\protect\citeauthoryear{{Patsourakos} et~al.}{2020}]{patsourakos2020}
\begin{barticle}
\bauthor{\bsnm{{Patsourakos}}, \binits{S.}},
\bauthor{\bsnm{{Vourlidas}}, \binits{A.}},
\bauthor{\bsnm{{T{\"o}r{\"o}k}}, \binits{T.}},
\bauthor{\bsnm{{Kliem}}, \binits{B.}},
\bauthor{\bsnm{{Antiochos}}, \binits{S.K.}},
\bauthor{\bsnm{{Archontis}}, \binits{V.}},
\bauthor{\bsnm{{Aulanier}}, \binits{G.}},
\bauthor{\bsnm{{Cheng}}, \binits{X.}},
\bauthor{\bsnm{{Chintzoglou}}, \binits{G.}},
\bauthor{\bsnm{{Georgoulis}}, \binits{M.K.}},
\bauthor{\bsnm{{Green}}, \binits{L.M.}},
\bauthor{\bsnm{{Leake}}, \binits{J.E.}},
\bauthor{\bsnm{{Moore}}, \binits{R.}},
\bauthor{\bsnm{{Nindos}}, \binits{A.}},
\bauthor{\bsnm{{Syntelis}}, \binits{P.}},
\bauthor{\bsnm{{Yardley}}, \binits{S.L.}},
\bauthor{\bsnm{{Yurchyshyn}}, \binits{V.}},
\bauthor{\bsnm{{Zhang}}, \binits{J.}}:
\byear{2020},
\batitle{{Decoding the Pre-Eruptive Magnetic Field Configurations of Coronal
  Mass Ejections}}.
\bjtitle{\ssr}
\bvolume{216},
\bfpage{131}.
\doiurl{https://doi.org/10.1007/s11214-020-00757-9}.
\adsurl{2020SSRv..216..131P}.
\end{barticle}
\endbibitem

\bibitem[\protect\citeauthoryear{{Pevtsov} et~al.}{2014}]{pevtsov2014}
\begin{barticle}
\bauthor{\bsnm{{Pevtsov}}, \binits{A.A.}},
\bauthor{\bsnm{{Berger}}, \binits{M.A.}},
\bauthor{\bsnm{{Nindos}}, \binits{A.}},
\bauthor{\bsnm{{Norton}}, \binits{A.A.}},
\bauthor{\bsnm{{van Driel-Gesztelyi}}, \binits{L.}}:
\byear{2014},
\batitle{{Magnetic Helicity, Tilt, and Twist}}.
\bjtitle{\ssr}
\bvolume{186},
\bfpage{285}.
\doiurl{https://doi.org/10.1007/s11214-014-0082-2}.
\adsurl{2014SSRv..186..285P}.
\end{barticle}
\endbibitem

\bibitem[\protect\citeauthoryear{{Poisson} et~al.}{2015}]{poisson2015}
\begin{barticle}
\bauthor{\bsnm{{Poisson}}, \binits{M.}},
\bauthor{\bsnm{{Mandrini}}, \binits{C.H.}},
\bauthor{\bsnm{{D{\'e}moulin}}, \binits{P.}},
\bauthor{\bsnm{{L{\'o}pez Fuentes}}, \binits{M.}}:
\byear{2015},
\batitle{{Evidence of Twisted Flux-Tube Emergence in Active Regions}}.
\bjtitle{\solphys}
\bvolume{290},
\bfpage{727}.
\doiurl{https://doi.org/10.1007/s11207-014-0633-4}.
\adsurl{2015SoPh..290..727P}.
\end{barticle}
\endbibitem

\bibitem[\protect\citeauthoryear{{Poisson} et~al.}{2020}]{poisson2020}
\begin{barticle}
\bauthor{\bsnm{{Poisson}}, \binits{M.}},
\bauthor{\bsnm{{Bustos}}, \binits{C.}},
\bauthor{\bsnm{{L{\'o}pez Fuentes}}, \binits{M.}},
\bauthor{\bsnm{{Mandrini}}, \binits{C.H.}},
\bauthor{\bsnm{{Cristiani}}, \binits{G.D.}}:
\byear{2020},
\batitle{{Two successive partial mini-filament confined ejections}}.
\bjtitle{Advances in Space Research}
\bvolume{65},
\bfpage{1629}.
\doiurl{https://doi.org/10.1016/j.asr.2019.09.026}.
\adsurl{2020AdSpR..65.1629P}.
\end{barticle}
\endbibitem

\bibitem[\protect\citeauthoryear{{Priest}, {Longcope}, and
  {Janvier}}{2016}]{priest2016}
\begin{barticle}
\bauthor{\bsnm{{Priest}}, \binits{E.R.}},
\bauthor{\bsnm{{Longcope}}, \binits{D.W.}},
\bauthor{\bsnm{{Janvier}}, \binits{M.}}:
\byear{2016},
\batitle{{Evolution of Magnetic Helicity During Eruptive Flares and Coronal
  Mass Ejections}}.
\bjtitle{\solphys}
\bvolume{291},
\bfpage{2017}.
\doiurl{https://doi.org/10.1007/s11207-016-0962-6}.
\adsurl{2016SoPh..291.2017P}.
\end{barticle}
\endbibitem

\bibitem[\protect\citeauthoryear{{Ruan} et~al.}{2014}]{ruan2014}
\begin{barticle}
\bauthor{\bsnm{{Ruan}}, \binits{G.}},
\bauthor{\bsnm{{Chen}}, \binits{Y.}},
\bauthor{\bsnm{{Wang}}, \binits{S.}},
\bauthor{\bsnm{{Zhang}}, \binits{H.}},
\bauthor{\bsnm{{Li}}, \binits{G.}},
\bauthor{\bsnm{{Jing}}, \binits{J.}},
\bauthor{\bsnm{{Su}}, \binits{J.}},
\bauthor{\bsnm{{Li}}, \binits{X.}},
\bauthor{\bsnm{{Xu}}, \binits{H.}},
\bauthor{\bsnm{{Du}}, \binits{G.}},
\bauthor{\bsnm{{Wang}}, \binits{H.}}:
\byear{2014},
\batitle{{A Solar Eruption Driven by Rapid Sunspot Rotation}}.
\bjtitle{\apj}
\bvolume{784},
\bfpage{165}.
\doiurl{https://doi.org/10.1088/0004-637X/784/2/165}.
\adsurl{2014ApJ...784..165R}.
\end{barticle}
\endbibitem

\bibitem[\protect\citeauthoryear{{Rust} and {Kumar}}{1994}]{rust1994}
\begin{barticle}
\bauthor{\bsnm{{Rust}}, \binits{D.M.}},
\bauthor{\bsnm{{Kumar}}, \binits{A.}}:
\byear{1994},
\batitle{{Helical Magnetic Fields in Filaments}}.
\bjtitle{\solphys}
\bvolume{155},
\bfpage{69}.
\doiurl{https://doi.org/10.1007/BF00670732}.
\adsurl{1994SoPh..155...69R}.
\end{barticle}
\endbibitem

\bibitem[\protect\citeauthoryear{{Sammis}, {Tang}, and
  {Zirin}}{2000}]{sammis2000}
\begin{barticle}
\bauthor{\bsnm{{Sammis}}, \binits{I.}},
\bauthor{\bsnm{{Tang}}, \binits{F.}},
\bauthor{\bsnm{{Zirin}}, \binits{H.}}:
\byear{2000},
\batitle{{The Dependence of Large Flare Occurrence on the Magnetic Structure of
  Sunspots}}.
\bjtitle{\apj}
\bvolume{540},
\bfpage{583}.
\doiurl{https://doi.org/10.1086/309303}.
\adsurl{2000ApJ...540..583S}.
\end{barticle}
\endbibitem

\bibitem[\protect\citeauthoryear{{Scherrer} et~al.}{1995}]{scherrer1995}
\begin{barticle}
\bauthor{\bsnm{{Scherrer}}, \binits{P.H.}},
\bauthor{\bsnm{{Bogart}}, \binits{R.S.}},
\bauthor{\bsnm{{Bush}}, \binits{R.I.}},
\bauthor{\bsnm{{Hoeksema}}, \binits{J.T.}},
\bauthor{\bsnm{{Kosovichev}}, \binits{A.G.}},
\bauthor{\bsnm{{Schou}}, \binits{J.}},
\bauthor{\bsnm{{Rosenberg}}, \binits{W.}},
\bauthor{\bsnm{{Springer}}, \binits{L.}},
\bauthor{\bsnm{{Tarbell}}, \binits{T.D.}},
\bauthor{\bsnm{{Title}}, \binits{A.}},
\bauthor{\bsnm{{Wolfson}}, \binits{C.J.}},
\bauthor{\bsnm{{Zayer}}, \binits{I.}},
\bauthor{\bsnm{{MDI Engineering Team}}}:
\byear{1995},
\batitle{{The Solar Oscillations Investigation - Michelson Doppler Imager}}.
\bjtitle{\solphys}
\bvolume{162},
\bfpage{129}.
\doiurl{https://doi.org/10.1007/BF00733429}.
\adsurl{1995SoPh..162..129S}.
\end{barticle}
\endbibitem

\bibitem[\protect\citeauthoryear{{Scherrer} et~al.}{2012}]{scherrer2012}
\begin{barticle}
\bauthor{\bsnm{{Scherrer}}, \binits{P.H.}},
\bauthor{\bsnm{{Schou}}, \binits{J.}},
\bauthor{\bsnm{{Bush}}, \binits{R.I.}},
\bauthor{\bsnm{{Kosovichev}}, \binits{A.G.}},
\bauthor{\bsnm{{Bogart}}, \binits{R.S.}},
\bauthor{\bsnm{{Hoeksema}}, \binits{J.T.}},
\bauthor{\bsnm{{Liu}}, \binits{Y.}},
\bauthor{\bsnm{{Duvall}}, \binits{T.L.}},
\bauthor{\bsnm{{Zhao}}, \binits{J.}},
\bauthor{\bsnm{{Title}}, \binits{A.M.}},
\bauthor{\bsnm{{Schrijver}}, \binits{C.J.}},
\bauthor{\bsnm{{Tarbell}}, \binits{T.D.}},
\bauthor{\bsnm{{Tomczyk}}, \binits{S.}}:
\byear{2012},
\batitle{{The Helioseismic and Magnetic Imager (HMI) Investigation for the
  Solar Dynamics Observatory (SDO)}}.
\bjtitle{\solphys}
\bvolume{275},
\bfpage{207}.
\doiurl{https://doi.org/10.1007/s11207-011-9834-2}.
\adsurl{2012SoPh..275..207S}.
\end{barticle}
\endbibitem

\bibitem[\protect\citeauthoryear{{Schrijver}}{2007}]{schrijver2007}
\begin{barticle}
\bauthor{\bsnm{{Schrijver}}, \binits{C.J.}}:
\byear{2007},
\batitle{{A Characteristic Magnetic Field Pattern Associated with All Major
  Solar Flares and Its Use in Flare Forecasting}}.
\bjtitle{\apjl}
\bvolume{655},
\bfpage{L117}.
\doiurl{https://doi.org/10.1086/511857}.
\adsurl{2007ApJ...655L.117S}.
\end{barticle}
\endbibitem

\bibitem[\protect\citeauthoryear{{Schuck}}{2008}]{schuck2008}
\begin{barticle}
\bauthor{\bsnm{{Schuck}}, \binits{P.W.}}:
\byear{2008},
\batitle{{Tracking Vector Magnetograms with the Magnetic Induction Equation}}.
\bjtitle{\apj}
\bvolume{683},
\bfpage{1134}.
\doiurl{https://doi.org/10.1086/589434}.
\adsurl{2008ApJ...683.1134S}.
\end{barticle}
\endbibitem

\bibitem[\protect\citeauthoryear{{Thalmann} et~al.}{2019}]{thalmann2019}
\begin{barticle}
\bauthor{\bsnm{{Thalmann}}, \binits{J.K.}},
\bauthor{\bsnm{{Moraitis}}, \binits{K.}},
\bauthor{\bsnm{{Linan}}, \binits{L.}},
\bauthor{\bsnm{{Pariat}}, \binits{E.}},
\bauthor{\bsnm{{Valori}}, \binits{G.}},
\bauthor{\bsnm{{Dalmasse}}, \binits{K.}}:
\byear{2019},
\batitle{{Magnetic Helicity Budget of Solar Active Regions Prolific of Eruptive
  and Confined Flares}}.
\bjtitle{\apj}
\bvolume{887},
\bfpage{64}.
\doiurl{https://doi.org/10.3847/1538-4357/ab4e15}.
\adsurl{2019ApJ...887...64T}.
\end{barticle}
\endbibitem

\bibitem[\protect\citeauthoryear{{Thalmann} et~al.}{2021}]{thalmann2021}
\begin{barticle}
\bauthor{\bsnm{{Thalmann}}, \binits{J.K.}},
\bauthor{\bsnm{{Georgoulis}}, \binits{M.K.}},
\bauthor{\bsnm{{Liu}}, \binits{Y.}},
\bauthor{\bsnm{{Pariat}}, \binits{E.}},
\bauthor{\bsnm{{Valori}}, \binits{G.}},
\bauthor{\bsnm{{Anfinogentov}}, \binits{S.}},
\bauthor{\bsnm{{Chen}}, \binits{F.}},
\bauthor{\bsnm{{Guo}}, \binits{Y.}},
\bauthor{\bsnm{{Moraitis}}, \binits{K.}},
\bauthor{\bsnm{{Yang}}, \binits{S.}},
\bauthor{\bsnm{{Mastrano}}, \binits{A.}},
\bauthor{\bsnm{{ISSI Team on Magnetic Helicity}}}:
\byear{2021},
\batitle{{Magnetic Helicity Estimations in Models and Observations of the Solar
  Magnetic Field. IV. Application to Solar Observations}}.
\bjtitle{\apj}
\bvolume{922},
\bfpage{41}.
\doiurl{https://doi.org/10.3847/1538-4357/ac1f93}.
\adsurl{2021ApJ...922...41T}.
\end{barticle}
\endbibitem

\bibitem[\protect\citeauthoryear{{Toriumi} and {Takasao}}{2017}]{toriumi2017b}
\begin{barticle}
\bauthor{\bsnm{{Toriumi}}, \binits{S.}},
\bauthor{\bsnm{{Takasao}}, \binits{S.}}:
\byear{2017},
\batitle{{Numerical Simulations of Flare-productive Active Regions:
  {\ensuremath{\delta}}-sunspots, Sheared Polarity Inversion Lines, Energy
  Storage, and Predictions}}.
\bjtitle{\apj}
\bvolume{850},
\bfpage{39}.
\doiurl{https://doi.org/10.3847/1538-4357/aa95c2}.
\adsurl{2017ApJ...850...39T}.
\end{barticle}
\endbibitem

\bibitem[\protect\citeauthoryear{{Toriumi} and {Wang}}{2019}]{toriumi2019}
\begin{barticle}
\bauthor{\bsnm{{Toriumi}}, \binits{S.}},
\bauthor{\bsnm{{Wang}}, \binits{H.}}:
\byear{2019},
\batitle{{Flare-productive active regions}}.
\bjtitle{Living Reviews in Solar Physics}
\bvolume{16},
\bfpage{3}.
\doiurl{https://doi.org/10.1007/s41116-019-0019-7}.
\adsurl{2019LRSP...16....3T}.
\end{barticle}
\endbibitem

\bibitem[\protect\citeauthoryear{{Toriumi} et~al.}{2017}]{toriumi2017}
\begin{barticle}
\bauthor{\bsnm{{Toriumi}}, \binits{S.}},
\bauthor{\bsnm{{Schrijver}}, \binits{C.J.}},
\bauthor{\bsnm{{Harra}}, \binits{L.K.}},
\bauthor{\bsnm{{Hudson}}, \binits{H.}},
\bauthor{\bsnm{{Nagashima}}, \binits{K.}}:
\byear{2017},
\batitle{{Magnetic Properties of Solar Active Regions That Govern Large Solar
  Flares and Eruptions}}.
\bjtitle{\apj}
\bvolume{834},
\bfpage{56}.
\doiurl{https://doi.org/10.3847/1538-4357/834/1/56}.
\adsurl{2017ApJ...834...56T}.
\end{barticle}
\endbibitem

\bibitem[\protect\citeauthoryear{{Tziotziou}, {Georgoulis}, and
  {Raouafi}}{2012}]{tziotziou2012}
\begin{barticle}
\bauthor{\bsnm{{Tziotziou}}, \binits{K.}},
\bauthor{\bsnm{{Georgoulis}}, \binits{M.K.}},
\bauthor{\bsnm{{Raouafi}}, \binits{N.-E.}}:
\byear{2012},
\batitle{{The Magnetic Energy-Helicity Diagram of Solar Active Regions}}.
\bjtitle{\apjl}
\bvolume{759},
\bfpage{L4}.
\doiurl{https://doi.org/10.1088/2041-8205/759/1/L4}.
\adsurl{2012ApJ...759L...4T}.
\end{barticle}
\endbibitem

\bibitem[\protect\citeauthoryear{{Tziotziou} et~al.}{2014}]{tziotziou2014}
\begin{barticle}
\bauthor{\bsnm{{Tziotziou}}, \binits{K.}},
\bauthor{\bsnm{{Moraitis}}, \binits{K.}},
\bauthor{\bsnm{{Georgoulis}}, \binits{M.K.}},
\bauthor{\bsnm{{Archontis}}, \binits{V.}}:
\byear{2014},
\batitle{{Validation of the magnetic energy vs. helicity scaling in solar
  magnetic structures}}.
\bjtitle{\aap}
\bvolume{570},
\bfpage{L1}.
\doiurl{https://doi.org/10.1051/0004-6361/201424864}.
\adsurl{2014A&A...570L...1T}.
\end{barticle}
\endbibitem

\bibitem[\protect\citeauthoryear{{Valori} et~al.}{2013}]{valori2013}
\begin{barticle}
\bauthor{\bsnm{{Valori}}, \binits{G.}},
\bauthor{\bsnm{{D{\'e}moulin}}, \binits{P.}},
\bauthor{\bsnm{{Pariat}}, \binits{E.}},
\bauthor{\bsnm{{Masson}}, \binits{S.}}:
\byear{2013},
\batitle{{Accuracy of magnetic energy computations}}.
\bjtitle{\aap}
\bvolume{553},
\bfpage{A38}.
\doiurl{https://doi.org/10.1051/0004-6361/201220982}.
\adsurl{2013A&A...553A..38V}.
\end{barticle}
\endbibitem

\bibitem[\protect\citeauthoryear{{Valori} et~al.}{2016}]{valori2016}
\begin{barticle}
\bauthor{\bsnm{{Valori}}, \binits{G.}},
\bauthor{\bsnm{{Pariat}}, \binits{E.}},
\bauthor{\bsnm{{Anfinogentov}}, \binits{S.}},
\bauthor{\bsnm{{Chen}}, \binits{F.}},
\bauthor{\bsnm{{Georgoulis}}, \binits{M.K.}},
\bauthor{\bsnm{{Guo}}, \binits{Y.}},
\bauthor{\bsnm{{Liu}}, \binits{Y.}},
\bauthor{\bsnm{{Moraitis}}, \binits{K.}},
\bauthor{\bsnm{{Thalmann}}, \binits{J.K.}},
\bauthor{\bsnm{{Yang}}, \binits{S.}}:
\byear{2016},
\batitle{{Magnetic Helicity Estimations in Models and Observations of the Solar
  Magnetic Field. Part I: Finite Volume Methods}}.
\bjtitle{\ssr}
\bvolume{201},
\bfpage{147}.
\doiurl{https://doi.org/10.1007/s11214-016-0299-3}.
\adsurl{2016SSRv..201..147V}.
\end{barticle}
\endbibitem

\bibitem[\protect\citeauthoryear{{van Driel-Gesztelyi} and
  {Green}}{2015}]{vandriel2015}
\begin{barticle}
\bauthor{\bsnm{{van Driel-Gesztelyi}}, \binits{L.}},
\bauthor{\bsnm{{Green}}, \binits{L.M.}}:
\byear{2015},
\batitle{{Evolution of Active Regions}}.
\bjtitle{Living Reviews in Solar Physics}
\bvolume{12},
\bfpage{1}.
\doiurl{https://doi.org/10.1007/lrsp-2015-1}.
\adsurl{2015LRSP...12....1V}.
\end{barticle}
\endbibitem

\bibitem[\protect\citeauthoryear{{Vemareddy}, {Cheng}, and
  {Ravindra}}{2016}]{vemareddy2016}
\begin{barticle}
\bauthor{\bsnm{{Vemareddy}}, \binits{P.}},
\bauthor{\bsnm{{Cheng}}, \binits{X.}},
\bauthor{\bsnm{{Ravindra}}, \binits{B.}}:
\byear{2016},
\batitle{{Sunspot Rotation as a Driver of Major Solar Eruptions in the NOAA
  Active Region 12158}}.
\bjtitle{\apj}
\bvolume{829},
\bfpage{24}.
\doiurl{https://doi.org/10.3847/0004-637X/829/1/24}.
\adsurl{2016ApJ...829...24V}.
\end{barticle}
\endbibitem

\bibitem[\protect\citeauthoryear{{Wang} et~al.}{2017}]{wang2017}
\begin{barticle}
\bauthor{\bsnm{{Wang}}, \binits{H.}},
\bauthor{\bsnm{{Liu}}, \binits{C.}},
\bauthor{\bsnm{{Ahn}}, \binits{K.}},
\bauthor{\bsnm{{Xu}}, \binits{Y.}},
\bauthor{\bsnm{{Jing}}, \binits{J.}},
\bauthor{\bsnm{{Deng}}, \binits{N.}},
\bauthor{\bsnm{{Huang}}, \binits{N.}},
\bauthor{\bsnm{{Liu}}, \binits{R.}},
\bauthor{\bsnm{{Kusano}}, \binits{K.}},
\bauthor{\bsnm{{Fleishman}}, \binits{G.D.}},
\bauthor{\bsnm{{Gary}}, \binits{D.E.}},
\bauthor{\bsnm{{Cao}}, \binits{W.}}:
\byear{2017},
\batitle{{High-resolution observations of flare precursors in the low solar
  atmosphere}}.
\bjtitle{Nature Astronomy}
\bvolume{1},
\bfpage{0085}.
\doiurl{https://doi.org/10.1038/s41550-017-0085}.
\adsurl{2017NatAs...1E..85W}.
\end{barticle}
\endbibitem

\bibitem[\protect\citeauthoryear{{Welsch} et~al.}{2009}]{welsch2009}
\begin{barticle}
\bauthor{\bsnm{{Welsch}}, \binits{B.T.}},
\bauthor{\bsnm{{Li}}, \binits{Y.}},
\bauthor{\bsnm{{Schuck}}, \binits{P.W.}},
\bauthor{\bsnm{{Fisher}}, \binits{G.H.}}:
\byear{2009},
\batitle{{What is the Relationship Between Photospheric Flow Fields and Solar
  Flares?}}
\bjtitle{\apj}
\bvolume{705},
\bfpage{821}.
\doiurl{https://doi.org/10.1088/0004-637X/705/1/821}.
\adsurl{2009ApJ...705..821W}.
\end{barticle}
\endbibitem

\bibitem[\protect\citeauthoryear{{Yang} and {Zhang}}{2018}]{yang2018}
\begin{barticle}
\bauthor{\bsnm{{Yang}}, \binits{S.}},
\bauthor{\bsnm{{Zhang}}, \binits{J.}}:
\byear{2018},
\batitle{{Mini-filament Eruptions Triggering Confined Solar Flares Observed by
  ONSET and SDO}}.
\bjtitle{\apjl}
\bvolume{860},
\bfpage{L25}.
\doiurl{https://doi.org/10.3847/2041-8213/aacaf9}.
\adsurl{2018ApJ...860L..25Y}.
\end{barticle}
\endbibitem

\bibitem[\protect\citeauthoryear{{Yang} et~al.}{2017}]{yang2017}
\begin{barticle}
\bauthor{\bsnm{{Yang}}, \binits{Y.-H.}},
\bauthor{\bsnm{{Hsieh}}, \binits{M.-S.}},
\bauthor{\bsnm{{Yu}}, \binits{H.-S.}},
\bauthor{\bsnm{{Chen}}, \binits{P.F.}}:
\byear{2017},
\batitle{{A Statistical Study of Flare Productivity Associated with Sunspot
  Properties in Different Magnetic Types of Active Regions}}.
\bjtitle{\apj}
\bvolume{834},
\bfpage{150}.
\doiurl{https://doi.org/10.3847/1538-4357/834/2/150}.
\adsurl{2017ApJ...834..150Y}.
\end{barticle}
\endbibitem

\bibitem[\protect\citeauthoryear{{Zhang}, {Liu}, and {Zhang}}{2008}]{zhang2008}
\begin{barticle}
\bauthor{\bsnm{{Zhang}}, \binits{Y.}},
\bauthor{\bsnm{{Liu}}, \binits{J.}},
\bauthor{\bsnm{{Zhang}}, \binits{H.}}:
\byear{2008},
\batitle{{Relationship between Rotating Sunspots and Flares}}.
\bjtitle{\solphys}
\bvolume{247},
\bfpage{39}.
\doiurl{https://doi.org/10.1007/s11207-007-9089-0}.
\adsurl{2008SoPh..247...39Z}.
\end{barticle}
\endbibitem

\bibitem[\protect\citeauthoryear{{Zuccarello} et~al.}{2018}]{zuccarello2018}
\begin{barticle}
\bauthor{\bsnm{{Zuccarello}}, \binits{F.P.}},
\bauthor{\bsnm{{Pariat}}, \binits{E.}},
\bauthor{\bsnm{{Valori}}, \binits{G.}},
\bauthor{\bsnm{{Linan}}, \binits{L.}}:
\byear{2018},
\batitle{{Threshold of Non-potential Magnetic Helicity Ratios at the Onset of
  Solar Eruptions}}.
\bjtitle{\apj}
\bvolume{863},
\bfpage{41}.
\doiurl{https://doi.org/10.3847/1538-4357/aacdfc}.
\adsurl{2018ApJ...863...41Z}.
\end{barticle}
\endbibitem

\end{thebibliography}

%
%
%
%

\end{article} 
\end{document}